\documentclass[amsmath,amssymb,12pt,eps,amsfonts]{article}
\usepackage{epsfig,epic}
\usepackage{amsfonts,amssymb,amsmath,epsfig,epic,color}
\usepackage{graphicx}
\usepackage{dcolumn}
\usepackage{bm}
\newcommand{\be}{\begin{equation}}
\newcommand{\ee}{\end{equation}}
\newcommand{\beqs}{\begin{eqnarray}}
\newcommand{\eeqs}{\end{eqnarray}}
\newcommand{\hbarm} { \frac{\hbar^2}{2m} }
\DeclareMathOperator*{\essup}{\textrm{Essup}}
\begin{document}
\title{ \bf Point interactions \\ in two and three dimensional Riemannian manifolds}

\author{\centerline {\small  Fatih Erman$^1$, O. Teoman Turgut$^{1,\,2}$}
\\\and
 {\scriptsize{$^1$
Department of Physics, Bo\u{g}azi\c{c}i University, Bebek, 34342,
\.Istanbul, Turkey}}
\\\and
{\scriptsize{$^2$Feza G\"{u}rsey Institute, Kuleli Mahallesi,
\c{S}ekip Ayhan \"{O}z{\i}\c{s}{\i}k Caddesi, No: 44, Kandilli,
34684, \.{I}stanbul, Turkey}}
\\
{\scriptsize{Electronic mail: fatih.erman@gmail.com,
turgutte@boun.edu.tr}}}

\date{\scriptsize{\textsc{\today}}}

\maketitle

%\date{\scriptsize{\textsc{\today}}}

Pacs Numbers: 11.10.Gh, 03.65.-w, 03.65.Ge

\abstract{We present a non-perturbative renormalization of the
bound state problem of $n$ bosons interacting  with finitely many
Dirac delta interactions on two and three dimensional Riemannian
manifolds using the heat kernel. We formulate the problem in terms
of a new operator called the principal or characteristic operator
$\Phi(E)$. In order to investigate the problem in more detail, we
then restrict the problem to one particle sector. The lower bound
of the ground state energy is found for general class of
manifolds, e.g., for compact and Cartan-Hadamard manifolds. The
estimate of the bound state energies in the tunneling regime is
calculated by perturbation theory. Non-degeneracy and uniqueness
of the ground state is proven by Perron-Frobenius theorem.
Moreover, the pointwise bounds on the wave function is given and
all these results are consistent with the one given in standard
quantum mechanics. Renormalization procedure does not lead to any
radical change in these cases. Finally, renormalization group
equations are derived and the $\beta$ function is exactly
calculated. This work is a natural continuation of our previous
work based on a novel approach to the renormalization of point
interactions, developed by S. G. Rajeev}.

\section{Introduction}

The studies of Dirac-delta interactions in quantum mechanics
(which are also called zero range, contact or point interactions,
or Fermi pseudopotentials in the literature) date back to the work
of Kronig and Penney \cite{kronig} who introduced the periodic
delta interactions describing the non-relativistic electrons
moving in a one dimensional fixed crystal lattice. The historical
development of point interactions has been given extensively in
the monograph \cite{Albeverio 2004}. Although it was Thomas
\cite{thomas} who pointed out that the problem of point
interactions in three dimensions could not be physically
acceptable due to the ultraviolet divergences, Thorn \cite{Thorn}
realized that we did not have to abandon these interactions and
physical results could be obtained after regularization and
renormalization, well known procedures in quantum field theory.

The motivation of studying point interactions in two and three
dimensions is based on trying to understand the concept of
renormalization in a simple context rather than field theory.
There are large amount of works on the renormalization of point
interactions in the literature from several point of views
\cite{thomas} - \cite{philips}: Regularization schemes can be
performed either in coordinate space \cite{Tarrach1, Tarrach2,
Mead, Perez} or in momentum space \cite{Thorn, huang, Jackiw,
philips, Mitra, henderson, Nyeo, Adhikari}. A single point
interaction in two dimensional flat space is also an instructive
example of dimensional transmutation \cite{Thorn, huang, Coleman,
Camblong}, that is, the original Hamiltonian does not contain any
intrinsic energy scale due to the dimensionless coupling constant
in natural units, but a new parameter specifying the bound state
energy, is introduced after the renormalization procedure, which
then fixes the energy scale of the system and this is called
dimensional transmutation. This implies a violation of $SO(2,1)$
symmetry of the scale invariant potential, so it is one of the
simplest examples of anomaly or quantum mechanical symmetry
breaking \cite{Jackiw}. Furthermore, renormalization group
equations of point interactions have been discussed in
\cite{Tarrach1,Adhikari,Camblong2} and the $\beta$ function has
been calculated exactly so that the theory has been found as
asymptotically free in two dimensions.

Mathematically rigorous treatments of point interactions are given
in the context of self-adjoint extension theory \cite{berezin
fadeev} and a detailed exposition of this subject has been
discussed in the monograph \cite{Albeverio 2004}. Dirac delta
interaction is considered as a self-adjoint extension of a
formally Hermitian free Hamiltonian on a space with a removed
point. The result of this method is identical to that of the
renormalization method if a certain relation between the parameter
in the extension and the renormalized (or bare) coupling constant
is satisfied \cite{Jackiw}. Self-adjoint extension method can be
also analyzed within the Green's function method \cite{Park,
Albeverio 2000}.

Many body version of this problem on $\mathbb{R}^2$ and
$\mathbb{R}^3$ is known as the formal non-relativistic limit of
the $\lambda \phi^4$ scalar field theory in (2+1) and (3+1)
dimensions. All these are extensively discussed first in the
unpublished thesis of J. Hoppe \cite{Hoppe} and later from a new
perspective in \cite{rajeevbound,rajeevdimock}. S. G. Rajeev
\cite{rajeevbound} introduced a new non-perturbative
renormalization method for point interactions which can be applied
to several many body theories: quantum mechanics with point
interactions, fermionic and bosonic quantum fields interacting
with a point source, many body problems with point interactions,
and non-relativistic field theory with polynomial interactions.
One of the main advantages of this approach is that all the
information about the spectrum of the model is described by an
explicit formula instead of imposing the boundary conditions on
the operator as in the case of self-adjoint extension theory. This
method is also particularly useful for dealing with the bound
state problems because of its non-perturbative nature. We are not
going to review the ideas developed in there. Instead, we suggest
the reader to read through the paper \cite{rajeevbound} to make
the reading of this paper easier.

Following the original ideas developed in \cite{rajeevbound}, we
previously  considered \textit{the bound state problem} for $N$
point interactions in two and three dimensional Riemannian
manifolds \cite{erman}. The lower bounds on the ground state
energy are found for few special manifolds $\mathbb{S}^2$,
$\mathbb{H}^2$ and $\mathbb{H}^3$ \cite{erman}. The construction
of the relativistic extension of this model in two dimensions is
also possible \cite{caglar}. We also applied the same method based
on the heat kernel to non-relativistic Lee model
\cite{nrleemodelonmanifold} and its relativistic version has been
constructed later on \cite{rleemodelonmanifold}. Our primary
motivation comes from the question, how the renormalization method
for the point interactions in quantum mechanics should be
performed non-perturbatively on Riemannian manifolds, hoping that
we can extend our understanding to the realm of quantum field
theory. It is worth pointing out that this problem on two
dimensional Riemannian manifolds also displays a kind of
dimensional transmutation, where new energy scales different from
the intrinsic energy scales of the system appear after the
renormalization \cite{erman}.

We organize this paper as follows. In section \ref{An Alternative
Construction of the Model}, we construct \textit{the bound state
problem} for $n$ bosons living in two and three dimensional
Riemannian manifolds interacting with $N$ external Dirac delta
interactions. This construction is motivated by the work
\cite{rajeevbound} in which the many body version of this problem
is renormalized non-perturbatively in flat spaces. We then
restrict the problem to $n=1$, and find the finite formulation of
the problem in terms of a new operator, which is called principal
operator (or characteristic operator) \cite{rajeevbound} and the
result is consistent with the one that we have obtained in
\cite{erman}. In section \ref{Interlacing Theorem and Perturbation
Theory}, we find the wave function for the bound states and show
that ground state energy of $N+1$ center case is smaller than the
$N$ center case using Cauchy interlacing theorem. Then, we
calculate the bound state energies in the tunnelling regime using
a version of perturbation theory. The results for our own purposes
on the upper and the lower bounds of heat kernel in the
mathematics literature is given shortly for compact and
Cartan-Hadamard manifolds in subsection \ref{heat kernel compact}
and in \ref{heat kernel noncompact}, respectively. Section
\ref{Pointwise Bounds on Wave function} presents pointwise bounds
on the wave function using the upper bounds of the heat kernel
given in the previous section and it is shown that the result is
consistent with the classical result found in the standard quantum
mechanics. Section \ref{lowerbound gse delta} establishes the
lower bound of the ground state energy for more general class of
Riemannian manifolds. The proof of the lower bound for the ground
state energy has the same spirit with the one given in our
previous work \cite{erman} but the proof given here is generalized
to a rich class of manifolds, such as compact and Cartan-Hadamard
manifolds. Then, in section \ref{Non-degeneracy and Positivity of
the Ground State}, non-degeneracy and positivity of the ground
state is proven with the help of Perron-Frobenius theorem.
Finally, we proceed with the study of the renormalization group
equations and the $\beta$ function is calculated exactly. Some
important properties and asymptotic expansion of the heat kernel
is summarized in the appendix A and the proof of the existence of
the Hamiltonian in two dimensions is explicitly given in the
appendix B.

\section{Construction of the Model}
\label{An Alternative Construction of the Model}

We consider $n$ nonrelativistic bosons living in a two or three
dimensional ($D=2,3$) Riemannian manifold and they interact with
$N$ external attractive Dirac delta potentials. In the second
quantized language, Hamiltonian of the system is
\be H=  \int_{\mathcal{M}} \mathrm{d}_{g}^{D} x \left[
\phi_{g}^{\dagger}(x) \left(-{\hbar^2 \over 2m}
\nabla_{g}^{2}\right) \phi_{g}(x) - \sum_{i=1}^{N} \lambda_i
\phi_{g}^{\dagger}(x) \delta_g(x,a_i) \phi_{g}(x) \right] \;, \ee
where $\mathrm{d}_{g}^{D} x = \sqrt{\det g}\, \mathrm{d}^{D} x $
is the $D$ dimensional volume element, $\nabla^2_g$ is
Laplace-Beltrami operator in the local coordinates $x\equiv(x^{1},
x^{2},\ldots, x^{D})$
\be \nabla^2_g = \frac{1}{\sqrt{\mathrm{det}\,g}}
\sum_{\alpha,\beta=1}^{D} \frac{\partial}{\partial x^\alpha}
\left(g^{\alpha \beta} \, \sqrt{\mathrm{det} \,g} \;
\frac{\partial}{\partial x^\beta}\right)\;, \ee
and $\phi^{\dag}_{g}(x)$, $\phi_{g}(x)$ is defined as the bosonic
creation-annihilation operators on the Riemannian manifold with
metric structure $g$. Here $a_i$ denotes the location of $i$-th
external dirac delta potential on the manifold and $\lambda_{i}
\in \mathbb{R}^+$ is the strength of the delta interaction at
$a_i$, so called coupling constant.

It is easy to show that the number of bosons $\int_{\mathcal{M}}
\mathrm{d}_{g}^{D} x \;\phi^{\dag}_{g}(x)\,\phi_{g}(x)$ is
conserved. The lowest eigenvalue of the Hamiltonian $H$ of our
problem in a sector with fixed number of bosons is either zero or
negative infinite, so the energy of the ground state is not
bounded from below \cite{erman}: $E \rightarrow -\infty$. The
first step we must do is to regularize the model. The natural
regularization of Hamiltonian can be chosen as
\beqs H_\epsilon = H_0 - \sum_{i=1}^{N} \lambda_i(\epsilon)
\int_{\mathcal{M}^2} \mathrm{d}_{g}^{D} x \, \mathrm{d}_{g}^{D} y
\; K_{\epsilon/2}(x,a_i;g) K_{\epsilon /2}(y,a_i;g)
\phi_{g}^{\dag}(y) \phi_{g}(x) \;, \eeqs
where $H_0$ is the free Hamiltonian. $K_{\epsilon} (x,y;g)$ is the
heat kernel defined on the Riemannian manifold $(\mathcal{M},g)$
and it converges to Dirac delta function $\delta_g(x,y)$ as
$\epsilon\rightarrow 0^+$. In this limit, one can see that we
recover the original Hamiltonian we are interested in. We write
the heat kernel as $K_{\epsilon} (x,y;g)$ throughout the paper in
order to specify which metric structure it is associated with.
Some essential properties of the heat kernel on Riemannian
manifolds that we have used in this paper are given in the
appendix A.

Now, we will consider the resolvent of the regularized Hamiltonian
in a Fock space formalism with arbitrary number of bosons.
Following the same methodology developed for the model in the
plane \cite{rajeevbound}, we shall extend the bosonic Fock space
$\mathcal{B}$ that we have started with, to $\mathcal{\tilde{B}} =
\mathcal{B} \oplus \mathcal{B} \otimes \mathbb{C}^N$ by defining
new creation and annihilation operators at the locations of the
dirac delta interactions. These are called \emph{angels}, which is
first introduced in \cite{rajeevbound}. The angel states allow us
rewrite the model in such a way that the coupling constant appears
additively rather than multiplicatively. As a result, we can
renormalize the model nonperturbatively by simply normal ordering.
We assume that the angel operators obey the orthofermionic algebra
\cite{mishra} defined by the following product relations (not with
commutators):
\beqs \chi_i \, \chi^{\dag}_j + \delta_{ij} \sum_{k=1}^{N}
\chi^{\dag}_k \, \chi_k
 = \mathbf{1} \delta_{ij},
\;\;\;\chi_i \, \chi_j = 0 = \chi^{\dag}_i \, \chi^{\dag}_j
\;,\eeqs
where $\mathbf{1}$ is the identity operator and $i,j,k=1,2,\ldots,
N$. It is more convenient for our purposes to write the angel
algebra in terms of projection operators:
\be \chi_i \, \chi^{\dag}_j =  \delta_{ij} \Pi_0, \;\;\;\chi_i \,
\chi_j=0= \chi^{\dag}_i \, \chi^{\dag}_j \;,\label{algebra of
angels} \ee
where
\be \Pi_1 = \sum_{k=1}^{N} \, \chi^{\dag}_k \, \chi_k ,
\;\;\;\Pi_0 = \mathbf{1}- \Pi_1 \;, \ee
are the projection operators onto the 1-angel and no-angel states,
respectively. Now we define the augmented regularized Hamiltonian
$\tilde{H}_{\epsilon}$ on $\mathcal{\tilde{B}}$
\beqs
 \tilde{H}_{\epsilon} = H_0 \Pi_0 + \bigg[\sum_{i=1}^{N} \int_{\mathcal{M}} \mathrm{d}_{g}^{D} x \;
 K_{\epsilon /2}(x,a_i ;g) \phi_g(x) \chi^{\dag}_i + h.c. \bigg]
 + \sum_{i=1}^{N} {1 \over \lambda_i(\epsilon)} \chi^{\dag}_i \, \chi_i
 \;.
\eeqs
If we split the Hilbert space according to the angel number, the
corresponding operator $\tilde{H}_{\epsilon} - E \Pi_0$ can be
written in the following matrix form:
\be
\tilde{H}_{\epsilon} - E \Pi_0 = \left(%
\begin{array}{cc}
  a & b^{\dagger}_{\epsilon} \\
  b_{\epsilon} & d_{\epsilon} \\
\end{array}%
\right)\;, \ee
with $a: \mathcal{B}\rightarrow \mathcal{B}$,
$b^{\dagger}_{\epsilon} : \mathcal{B} \otimes \mathbb{C}^N
\rightarrow \mathcal{B}$, $d_{\epsilon}: \mathcal{B} \otimes
\mathbb{C}^N \rightarrow \mathcal{B} \otimes \mathbb{C}^N $. Here,
\beqs a&=& H_0 - E \;, \hspace{2cm} d_{\epsilon}=\sum_{i=1}^{N} {1
\over \lambda_i(\epsilon)} \chi^{\dag}_i \, \chi_i  \cr
b^{\dag}_{\epsilon}&=& \sum_{i=1}^{N} \int_{\mathcal{M}}
\mathrm{d}_{g}^{D} x \;
 K_{\epsilon/2}(x,a_i ;g) \phi_{g}^{\dag}(x)\, \chi_i \;. \eeqs
Then, one can construct the augmented regularized resolvent
\be
\tilde{R_{\epsilon}}(E)= {1 \over \tilde{H}_{\epsilon} -E \Pi_0}= \left(%
\begin{array}{cc}
  \alpha_{\epsilon} & \beta^\dagger_{\epsilon} \\
  \beta_{\epsilon} & \delta_{\epsilon} \\
\end{array}%
\right)\;. \ee
One can find
$\alpha_{\epsilon},\beta_{\epsilon},\delta_{\epsilon}$ in terms of
$a,b_{\epsilon},d_{\epsilon}$ by direct computation. This could be
done apparently different but equivalent ways and the formulas
were obtained in the appendix of \cite{rajeevbound}
\be \alpha_{\epsilon} = \left[a- b^\dag_{\epsilon}
\,d^{-1}_{\epsilon} \, b_{\epsilon} \right]^{-1}={1 \over
H_{\epsilon} -E} = R_{\epsilon}(E) \;.\ee
This means that $\tilde{R}_{\epsilon}(E)$ projected to
$\mathcal{B}$ is just the resolvent of the operator $H_\epsilon$.
We have also another formula for $\alpha_{\epsilon}$
\cite{rajeevbound}
\be \alpha_{\epsilon}  = a^{-1} +  a^{-1} \, b^\dag_{\epsilon}
\left[d_{\epsilon} -  b_{\epsilon} \, a^{-1} \, b^\dag_{\epsilon}
\right]^{-1} b_{\epsilon} \, a^{-1} \;,  \ee
or
\be R_{\epsilon}(E) = {1 \over H_0 -E} +  {1 \over H_0 -E} \,
b^\dag_{\epsilon} \Phi_{\epsilon}(E)^{-1} b_{\epsilon} \, {1 \over
H_0 -E} \;, \ee
where
\beqs & & \hskip-1cm \Phi_{\epsilon}(E) = \sum_{i=1}^{N} {1 \over
\lambda_i(\epsilon)} \chi^{\dag}_i \, \chi_i - \sum_{i,j=1}^{N}
\int_{\mathcal{M}^2} \mathrm{d}_{g}^{D} x \, \mathrm{d}_{g}^{D} y
\; K_{\epsilon /2}(x,a_i;g) K_{\epsilon /2}(y,a_j;g) \cr & \ &
\phi_{g}(y) \left( {1 \over H_0 -E} \right) \phi_{g}^{\dag}(x) \,
\chi^{\dag}_i \, \chi_j \label{Phiepsilon} \;. \eeqs
The operator $\Phi_{\epsilon}(E)$ is called the regularized
principal operator (or regularized characteristic operator
\cite{rajeevbound}). Note that writing the resolvent of
$H_{\epsilon}$ in this way allows us to write the coupling
constant additively. The renormalization procedure can then be
done if we can separate the singular part of the operator
$\Phi_\epsilon$. We will see that this is possible by normal
ordering of the operators in the principal operator. By using
eigenfunction expansions (\ref{expheat}) and (\ref{eigenfuncexp})
of the operators $\phi_{g}(x)$, $\phi^{\dag}_{g}(x)$ and that of
the heat kernel (or their analogs for non-compact manifolds), one
can shift the operator $\phi^{\dag}_{g}(x)$ in (\ref{Phiepsilon})
to the left
\be {1\over H_0-E} \phi^{\dag}_{g}(x)= \int_{\mathcal{M}}
\mathrm{d}_{g}^{D} x' \; \phi^{\dag}_{g}(x') \int_0^\infty
{\mathrm{d} t \over \hbar} \; e^{-{t \over \hbar}(H_0-E)} \,
K_{t}(x,x';g) \;, \ee
and one can also shift the operator $\phi_{g}(x)$ to the right
using a similar equation. Then, the normal ordered principal
operator can be written by using the properties of heat kernel and
separating the $i=j$ term from the sum
\beqs & & \hskip-1cm \Phi_{\epsilon}(E) = \sum_{i=1}^{N} {1 \over
\lambda_i(\epsilon)} \, \chi^{\dag}_i \, \chi_i - \sum_{i,j=1}^{N}
\int_{\mathcal{M}^2} \mathrm{d}_{g}^{D} x \, \mathrm{d}_{g}^{D} y
\; \int_{0}^{\infty} {\mathrm{d} t \over \hbar} K_{(t+
\epsilon/2)}(x,a_i;g) K_{(t+ \epsilon/2)}(y,a_j;g)
 \cr & & \phi_{g}^{\dag}(x) e^{-{t \over \hbar}(H_0 -E)}\phi_{g}(y)
\chi^{\dag}_i \, \chi_j - \sum_{i=1}^{N} \int_{0}^{\infty}
{\mathrm{d} t \over \hbar} K_{(\epsilon+t)}(a_i,a_i;g)  e^{-{t
\over \hbar}(H_0 -E)}
\chi^{\dag}_{i} \, \chi_{i} \cr & & - \sum_{\substack{i=1 \\
i \neq j}}^{N} \int_{0}^{\infty} {\mathrm{d} t \over \hbar}
K_{(\epsilon+t)}(a_i,a_j;g)  e^{-{t \over \hbar}(H_0 -E)}
\chi^{\dag}_{i} \, \chi_{j} \;. \eeqs
Due to the singular behavior of the diagonal heat kernel
(\ref{asymheat}) near $t=0$, we expect that the third term is
divergent as $\epsilon \rightarrow 0^+$. Therefore, if we choose
the coupling constant
\be {1 \over \lambda_i (\epsilon)} = \int_{\epsilon}^{\infty}
{\mathrm{d} t \over \hbar} K_{t}(a_i,a_i;g) e^{-{t \over
\hbar}\mu_{i}^2} \;, \label{coupling const renormalization} \ee
where $-\mu_i^2$ corresponds to experimentally measured bound
state energy of the individual $i$th Dirac delta center, we find
the principal operator after taking the limit $\epsilon
\rightarrow 0^+$
\beqs & & \Phi(E) = \lim_{\epsilon\rightarrow 0^+}
\Phi_{\epsilon}(E)=
 \sum_{i=1}^{N} \int_{0}^{\infty} {\mathrm{d} t \over \hbar}
K_{t} (a_i,a_i;g) \left( e^{-{t \over \hbar} \mu_{i}^2 } - e^{- {t
\over \hbar} (H_0 -E)} \right) \chi^{\dag}_i \, \chi_i \cr & \ & -
\sum_{i,j=1}^{N} \int_{0}^{\infty} {\mathrm{d} t \over \hbar}
\int_{\mathcal{M}^2} \mathrm{d}_{g}^{D} x \, \mathrm{d}_{g}^{D} y
\; K_{t} (a_i,x;g) K_{t} (a_j,y;g) \phi_{g}^{\dag}(x)
 e^{-{t \over \hbar}(H_0 -E)} \phi_{g}(y)\chi^{\dag}_i \, \chi_j
\cr & \ & - \sum_{\substack{i=1 \\
i \neq j}}^{N} \int_{0}^{\infty} {\mathrm{d} t \over \hbar}
K_{t}(a_i,a_j;g) e^{-{t \over \hbar}(H_0 -E)} \chi^{\dag}_i \,
\chi_j \label{phimanyparticledelta} \;. \eeqs
This can be written in a more compact way
$\Phi(E)=\sum_{i,j=1}^{N}\Phi_{ij}(E) \, \chi^{\dag}_i \, \chi_j
$, where $\Phi_{ij}(E)$ can be read from
(\ref{phimanyparticledelta}). Once we have a proper definition of
the principal operator, the divergence is completely removed since
the spectrum of the problem can be found from the resolvent. We
are now in a position to get the full resolvent of our problem in
terms of the principal operator
\beqs & & \hskip-1cm R(E) = \lim_{\epsilon \rightarrow 0^+}
R_{\epsilon}(E) = {1 \over H_0 -E} + {1 \over H_0 -E} \,
\sum_{k=1}^{N} \phi_{g}^{\dag}(a_k)\, \chi_k \, \Phi^{-1}(E)
\,\sum_{l=1}^{N} \phi_{g}(a_l) \chi^{\dag}_l  \, {1 \over H_0 -E}
\cr & & = {1 \over H_0 -E} + {1 \over H_0 -E} \, \sum_{i,j=1}^{N}
\phi_{g}^{\dag}(a_i) \, \Phi^{-1}_{ij}(E) \phi_{g}(a_j) \,
 {1 \over H_0 -E} \label{resolvent}  \;,\eeqs
where we have used $\Phi^{-1}(E)= \sum_{i,j=1}^{N}
\Phi^{-1}_{ij}(E) \, \chi^{\dag}_i \, \chi_j$ and the algebra of
angel operators (\ref{algebra of angels}) with the fact that
$R(E): \mathcal{B}\rightarrow \mathcal{B}$. We note that the
principal operator can be extended to its largest domain of
definition in the complex energy plane by analytic continuation.

Since $R(E): \mathcal{B}\rightarrow \mathcal{B}$, we can consider
the resolvent kernel between $n$ bosons with no angel states. Up
to here, we have generalized the construction of the problem given
in \cite{erman} to the many boson cases. From now on, we shall
study one boson problem for simplicity, that is,
\be \int_{\mathcal{M}} \mathrm{d}_{g}^{D} x \; \psi(x)
\phi_{g}^{\dag}(x) |0 \rangle \otimes |\Omega \rangle \;, \ee
where $|\Omega \rangle$ is the vacuum for the angel state and
$\psi(x)$ is the wave function for the boson. Then, the resolvent
kernel corresponding to this state satisfies the following
equation after a straightforward calculation
\be R(x,y|E)= R_0(x,y|E) + \sum_{i,j=1}^{N} R_0(x,a_i|E)
\Phi^{-1}_{ij}(E) R_0(a_j,y|E) \label{resolventkernel} \;,  \ee
where $R_0(x,y|E)$ is the free resolvent kernel which can be
written in terms of heat kernel (\ref{heatkernel and A}). Here
$\Phi_{ij}(E)$ can be analytically continued to its largest set in
the entire complex plane so $E$ should be considered as a complex
variable. The equation (\ref{resolventkernel}) gives the relation
between the resolvent defined on an infinite dimensional space and
the principal matrix defined on a finite dimensional space. Such
formulae were extensively discussed in problems associated with
self-adjoint extensions of operators, notably by Krein and his
school, and also for such singular interactions in flat spaces
\cite{Albeverio 2004, Jackiw, Albeverio 2000}. Therefore, our
problem can be considered also a kind of self-adjoint extension of
the free Hamiltonian. We will come back to this point at the end
of the section \ref{Pointwise Bounds on Wave function}. The
resolvent essentially includes all the information about the
spectrum. We will restrict ourselves only to the bound state
problem since the scattering problem requires a deeper analysis.
The discrete spectrum of the Hamiltonian is the set of numbers $E$
at which the resolvent does not exist and continuous spectrum
corresponds to the unbounded resolvent. The poles corresponding to
bound states must be due to the $\Phi^{-1}(E)$ which can be seen
from (\ref{resolventkernel}). In other words, the roots of
\be \Phi(E)|\Psi \rangle = 0 \;, \label{phi psi = zero} \ee
determine the bound state spectrum of the model. Since $\Phi(E):
\mathcal{B} \otimes \mathbb{C}^N \rightarrow \mathcal{B} \otimes
\mathbb{C}^N$, let us try to consider $|\Psi \rangle $ as a direct
product of no-boson with one angel state:
\be
 | \Psi \rangle  = |0 \rangle \otimes \sum_{k=1}^{N} A_k
|e_k \rangle \;, \ee
where $|e_k \rangle \equiv  \chi^{\dag}_k | \Omega \rangle $ is a
set of complete orthonormal basis for $\mathbb{C}^{N}$. Then, the
equation (\ref{phi psi = zero}) yields
\beqs  \sum_{i=1}^{N} \int_{0}^{\infty} {\mathrm{d} t \over \hbar}
K_{t} (a_i,a_i;g) \left( e^{-{t \over \hbar} \mu_{i}^2 } - e^{{t
\over \hbar} E}
\right) A_i |e_i \rangle - \sum_{\substack{i=1 \\
i \neq j}}^{N} \int_{0}^{\infty} {\mathrm{d} t \over \hbar}
K_{t}(a_i,a_j;g) e^{{t \over \hbar} E} A_j | e_i \rangle =0 \;,
\eeqs
and the result can be written as a matrix equation
\be \sum_{j=1}^{N} \Phi_{ij}(E)A_j =0 \;, \ee
where
\be  \Phi_{ij} (E) =
\begin{cases}
\begin{split}
\int_0 ^\infty {\mathrm{d} t \over \hbar} K_{t}(a_i,a_i;g)\left(
e^{-{t \over \hbar} \mu_i^2} - e^{{t \over \hbar} E} \right)
\end{split}
& \textrm{if $i = j$} \\
\begin{split}
- \; \int_0^\infty {\mathrm{d} t \over \hbar} K_{
t}(a_i,a_j;g)e^{{t \over \hbar} E}
\end{split}
& \textrm{if $i \neq j$}.
\end{cases} \label{principal operator}
\ee
Thanks to the symmetry property of the heat kernel
(\ref{symmetryprop}), the principal matrix is Hermitian for real
values of energy and the explicit form of it in terms of heat
kernel has been first obtained by a different method in our
previous work \cite{erman}. Many important aspects of the model
can be understood by working out the principal matrix as we will
see in the following sections.

It is well known that the same problem with a single delta
potential in flat spaces is a good example of a dimensional
transmutation in quantum mechanics. Our problem in $D=2$ realizes
a generalized dimensional transmutation \cite{Camblong}: In our
case, the coupling constants $\lambda_i$ have the same dimension
as ${\hbar^2 \over m}$ by dimensional analysis. In contrast to the
flat case, we also have intrinsic scales coming from the geometry
of the space, such as curvature and the geodesic distance between
centers $d_{ij}$. However, after the renormalization procedure, we
obtain a set of new dimensional parameters $\mu_i^2$ from the
relation (\ref{coupling const renormalization}). Hence, the energy
is not determined by naive dimensional analysis. However, in the
case of single delta attractor for the flat case there is no
combination of dimensional parameters to come up with an energy
scale, whereas in the case of a manifold we have geometric length
scales which already may define an energy scale. The dimensional
transmutation is most striking in cases where there is no
intrinsic energy scale.

\section{Interlacing Theorem and Perturbation Theory}
\label{Interlacing Theorem and Perturbation Theory}

A mathematically satisfactory calculation of the wave function
should proceed from the resolvent equation. Since the eigenvalues
are isolated we can find the projection operator to the subspace
corresponding to this eigenvalue by a contour integral
\cite{simon}:
\be \langle x| \mathbb{P}_{k}|y \rangle =
\psi_k(x)\psi_{k}^{*}(y)=- {1\over 2\pi i} \oint_{\Gamma_k}
\mathrm{d} z \;  R(x,y|z), \ee
where $\Gamma_k$ is a small contour enclosing the  isolated
eigenvalue $-\nu_k^2$. We note that the free Green's functions
$R_0(x,y|z)$ will not contain any poles on the negative real axis,
so all the poles on the negative real axis will come from the
poles of inverse principal matrix $\Phi^{-1}(z)$. For simplicity
we can assume that the eigenvalues are  nondegenerate and let us
denote the $k$ th eigenvalue of the principal matrix as $\omega^k$
so eigenvalue problem is
\be \label{eigenvalue principal delta} \sum_{j=1}^{N}
\Phi_{ij}(-\nu_k^2) A_j^{k}(-\nu_k^2)=\omega^k(-\nu_k^2)
A_j^{k}(-\nu_k^2)\;. \ee
Since the principal matrix is Hermitian on the real line and
\be \Phi^{\dagger}_{ij}(z)=\Phi_{ij}(z^{*})\;, \ee
on the complex plane, there exists a holomorphic family of
projection operators on the complex plane \cite{kato}, so that we
can apply the spectral theorem for the principal matrix $\Phi(z)$
\be \Phi_{ij}(z)=\sum_{k} \omega^k(z)\mathbb{P}_k(z)_{ij} \;, \ee
here $\mathbb{P}_k(z)_{ij}=A_i^{k*}(z)A_j^k(z)$, $A^k_i(z)$ is the
normalized eigenvector corresponding to the eigenvalue
$\omega^k(z)$. Similarly, we can write the spectral resolution of
the inverse principal matrix,
\be
 \Phi^{-1}_{ij}(z)=\sum_{k} {1 \over \omega^k(z)} \mathbb{P}_k(z)_{ij}
\;. \ee
The residue can then be found
\beqs & & \lim_{z \to -\nu^2_k} R_0(x,a_i|z) (z+\nu^{2}_{k})
\Phi^{-1}_{ij}(z) R_0(a_j,y|z) \cr & & = R_0(x,a_i|-\nu_k^2)\left[
\left.{\partial \omega^k(z)\over \partial
z}\right|_{z=-\nu_k^2}\right]^{-1}
 \mathbb{P}_k(-\nu_k^2)_{ij}R_0(a_j,y|-\nu_k^2)\;. \eeqs
Now we will look at the variations of the eigenvalues of $\Phi$ as
we change the parameters $\nu$. Using
\be
\omega^k(-\nu^2)=\left(A^{k}(-\nu^2),\Phi(-\nu^2)A^{k}(-\nu^2)\right)
\;,\ee
and as a consequence of Feynman-Hellman theorem \cite{feynman
hellmann} in the non-degenerate case, we have
\beqs
   {\partial \omega^k(-\nu^2)\over \partial\nu}&=&\left(A^{k}(-\nu^2),{\partial\Phi(-\nu^2)
\over \partial \nu} A^{k}(-\nu^2)\right) \cr &=& \sum_{i,j=1}^{N}
A^{k *}_{i}(-\nu^2) {\partial \Phi_{ij}(-\nu^2) \over \partial
\nu} A^{k}_{j}(-\nu^2) \;. \label{feynmanhelmann} \eeqs
Taking the derivative of the principal matrix with respect to
$\nu$ from (\ref{principal operator})
\be \left.{\partial \Phi_{ij}(-\nu^2) \over \partial
\nu}\right|_{\nu=\nu_{k}}= \int_{0}^{\infty} {\mathrm{d} t \over
\hbar} \; \left({2 \nu_k t \over \hbar} \right) K_t(a_i,a_j;g)
e^{-\frac{t \nu_{k}^2}{\hbar}} \;, \label{derivative of phi} \ee
and inserting the equation (\ref{derivative of phi}) into the
equation (\ref{feynmanhelmann}), we obtain
\beqs & & {\partial \omega^k(z)\over \partial z} =  - {1 \over 2
\nu}{\partial \omega^k(-\nu^2) \over \partial \nu} \cr
  & & \hskip-1cm = - {1 \over 2 \nu} \sum_{i,j=1}^{N} A^{k
*}_{i}(-\nu^2) \int_{0}^{\infty} {\mathrm{d} t \over \hbar} \;
\left({2 \nu t \over \hbar} \right) K_t(a_i,a_j;g) e^{-\frac{t
\nu^2}{\hbar}} A^{k}_{j}(-\nu^2) \;. \label{derivative of lambda}
\eeqs
If we evaluate (\ref{derivative of lambda}) at $z=-\nu_{k}^{2}$,
or at $\nu=\nu_k$, it yields
\beqs \left. {\partial \omega^k(z) \over \partial z}
\right|_{z=-\nu_k^2} = - \sum_{i,j=1}^{N} A^{k *}_{i}(-\nu_k^2)
\int_{0}^{\infty} {\mathrm{d} t \over \hbar} \; \left({t \over
\hbar} \right) K_t(a_i,a_j;g) e^{-\frac{t \nu_k^2}{\hbar}}
A^{k}_{j}(-\nu_k^2)\;. \label{derivative of lambda evaluated}
\eeqs
Note that the integral is finite in two and three dimensions due
to upper bounds on the heat kernel. If we combine all these
results, we get
\beqs \psi_k(x)\psi_{k}^{*}(y) &=& -{1\over 2 \pi i} (2 \pi i)
R_0(x,a_i|-\nu_k^2) \Bigg[ - \sum_{i,j=1}^{N} A^{*}_{i}(-\nu_k^2)
\int_{0}^{\infty} {\mathrm{d} t \over \hbar} \; \left({t \over
\hbar} \right)  \cr & & \hskip-2cm \times K_t(a_i,a_j;g)
e^{-\frac{t \nu_k^2}{\hbar}} A_{j}(-\nu_k^2) \Bigg]^{-1}
A_i(-\nu_k^2)^*A_j(-\nu_k^2) R_0(a_j,y|-\nu_k^2) .\eeqs
Then, we can directly read off the bound state wave function from
the equation above:
\be
\begin{split}
\psi_k(x) = & \left[ \sum_{i,j=1}^N A_i^{*}(-\nu_k^2)
\int_{0}^{\infty} {\mathrm{d} t \over \hbar} \; \left({t \over
\hbar} \right) K_t(a_i,a_j;g) e^{-\frac{t \nu_{k}^2}{\hbar}} \,
A_j(-\nu_k^2) \right]^{-\frac{1}{2}} \; \cr & \hspace{3cm}\times
\int_{0}^{\infty} {\mathrm{d} t  \over \hbar}\; e^{-\frac{t
\nu_{k}^2}{\hbar}} \sum_{i=1}^N A_i(-\nu_k^2) K_{t}(a_i, x;g) \;,
\label{wavefunction heat delta} \end{split} \ee
from which one can easily see that $\psi_k(x)$ is finite except at
$x=a_i$. This is exactly the same result obtained by a different
method in \cite{erman}. Note that we have written
$A_i^k(-\nu_k^2)$ as $A_i(-\nu_k^2)$ for simplicity. Incidentally
the equation (\ref{derivative of lambda evaluated}) implies an
interesting result for the variation of eigenvalues,
\beqs & & \left. {\partial \omega^k(-\nu^2) \over \partial \nu}
\right|_{\nu=\nu_k} = \sum_{i,j=1}^{N} A^{*}_{i}(-\nu_k^2)
\int_{0}^{\infty} {\mathrm{d} t \over \hbar} \; \left({2 \nu_k t
\over \hbar} \right) K_t(a_i,a_j;g) e^{-\frac{t \nu_k^2}{\hbar}}
A_{j}(-\nu_k^2) \cr  & &  = \sum_{i,j=1}^{N} A^{*}_{i}(-\nu_k^2)
\int_{0}^{\infty} {\mathrm{d} t \over \hbar} \; \left({2 \nu_k t
\over \hbar} \right) \int_{\mathcal{M}} \mathrm{d}_g x \;
K_{t/2}(a_i,x;g) K_{t/2}(x,a_j;g) e^{-\frac{t \nu_k^2}{\hbar}}
A_{j}(-\nu_k^2) \cr & & =   \int_{0}^{\infty} {\mathrm{d} t \over
\hbar} \; \left({2 \nu_k t \over \hbar} \right)  e^{-\frac{t
\nu_k^2}{\hbar}} \int_{\mathcal{M}} \mathrm{d}_g x \;
\sum_{i=1}^{N} \left|K_{t/2}(a_i,x;g) A_{i}(-\nu_k^2)\right|^2 \;,
\label{derivative of lambda evaluated nu} \eeqs
where we have used the fact that $\nu_k \in \mathbb{R}^+$ for all
$k$ and properties of heat kernel (\ref{semigroupprop}) and
(\ref{symmetryprop}) and the order of the integration and the
finite sum can be interchanged. We can easily see that the above
equation (\ref{derivative of lambda evaluated nu}) is strictly
positive due to the positivity of heat kernel, so
\be \left. {\partial \omega^k(-\nu^2) \over \partial \nu}
\right|_{\nu=\nu_k} > 0 \;. \label{derivative of lambda positive}
\ee
Energy eigenvalues $E=-\nu^2$ are obtained from the zeros of the
eigenvalues of the principal matrix, that is,
$\omega^k(-\nu_k^2)=0$, and there is a unique solution for each
$\omega^k(-\nu^2)$. We also know that for sufficiently small
values of $\nu$, the matrix $\Phi(-\nu^2)$ becomes negative, hence
no zeros exists beyond some critical point.

With this result in mind, let us see what can be said about the
comparison of the energy eigenvalues for the different number of
delta centers. In order to see this, we need the Cauchy
interlacing theorem in mathematics literature \cite{bhatia}, which
states that if we delete the last row and column  of an Hermitian
$(N+1)\times (N+1)$ matrix, the eigenvalues of the original matrix
is interlaced by the eigenvalues of the new matrix, i.e. if
$\omega^1 (-\nu^2) \leq \omega^2(-\nu^2) \leq \cdots \leq
\omega^{N+1}(-\nu^2)$ lists the eigenvalues of the original
$(N+1)\times (N+1)$ matrix and if $\tilde{\omega}^1 (-\nu^2) \leq
\tilde{\omega}^2(-\nu^2) \leq \cdots \leq
\tilde{\omega}^{N}(-\nu^2)$ lists the eigenvalues of the reduced
matrix (any $N \times N$ principal submatrix of the $(N+1)\times
(N+1)$ matrix), then we have
\be
 \omega^1(-\nu^2) \leq \tilde \omega^1(-\nu^2) \leq \omega^2(-\nu^2) \leq \tilde \omega^2(-\nu^2) \leq  \cdots
 \leq \tilde \omega^N(-\nu^2) \leq \omega^{N+1}(-\nu^2) \;. \ee
We assume that $(N+1)\times (N+1)$ matrix in the above-mentioned
theorem is the principal matrix $\Phi^{N+1}(-\nu^2)$ corresponding
to a certain arrangement of $N+1$ delta potentials. If we now
delete the last row and the column it means that we remove the
$(N+1)$th delta center from the system. The bound state problem of
$N$ centers corresponds to zero eigenvalue of the principal matrix
$\Phi^{N}$, and let us denote that bound state energy as
$\tilde{E}$:
\be \tilde \omega^k(- \tilde{\nu}_{k}^{* 2})=0, \ \ \
\tilde{E}_k=- \tilde{\nu}_{k}^{* 2} \;,\ee
if it exists. By the Cauchy interlacing result, we then expect the
following inequality
\be \cdots < \omega^k (-\tilde{\nu}_{k}^{* 2}) < \tilde
\omega^k(-\tilde{\nu}^{* 2}_{k})=0 <
\omega^{k+1}(-\tilde{\nu}_{k}^{* 2}) < \cdots \;, \ee
From the positivity of the derivative of the eigenvalues with
respect to the argument (\ref{derivative of lambda positive}), the
eigenvalues $\omega$ are monotonically increasing functions.
Hence, in order to get a zero root of $\omega^k
(-\tilde{\nu}_{k}^{* 2})$, we should increase $\tilde{\nu}_k^{*}$
to a higher value $\nu_k^{*}$. As a result, $\omega^k(-\nu^{*
2}_k)=0$ if $\nu_k^{* 2}
> \tilde \nu_k^{* 2}$, i.e. $E_k = - \nu_k^{*2}<\tilde E_k=-\tilde
\nu_k^{*2}$. Thus, the energies also interlace in the same
manner--this is a nonlinear
 analog of Sturm's comparison theorem of the eigenvalues.
Moreover, $E_{gr}^{N+1} < E_{gr}^N<E_{gr}^1=-\mu_k^2$, that is,
the ground state is always negative and approaches the bound state
energy for the one delta center as $N$ gets smaller. We can also
generalize these results to the degenerate cases but the proof is
more cumbersome.

It is worth noting that we do not have to solve the energy
eigenvalues for the bound state while performing our
non-perturbative renormalization method. In other words, although
our renormalization method makes the problem well defined and
finite, the energy eigenvalues must be found after this finite
formulation has been constructed. If we cannot solve the problem
exactly after the renormalization, we must apply the standard
approximation methods, such as perturbation theory and variational
techniques. An interesting estimate for our problem can be given
by perturbation theory. For simplicity, we assume that all binding
energies $-\mu_k^2$ s are different and the magnitude of the
minimum binding energy of the $k$-th singular potential is much
larger than the correlation energy between the $k$-th fixed center
and the $l$-th center, that is to say, we assume ${\hbar^2\over 2m
d^2(a_k,a_l)} \ll \mu_{k, min}^{2}$ (on a noncompact manifold we
may assume that the geodesic distance between the centers is
large). This assumption makes the off-diagonal elements of the
principal matrix much smaller than its diagonal elements. For this
reason, let us separate the principal matrix for $E_k=-\nu_k^2$ as
the sum of a diagonal matrix and an off-diagonal matrix, which is
very small compared to the diagonal part:
\be \Phi(-\nu_k^2)=\Phi_D(-\nu_k^2)+ \delta \Phi(-\nu_k^2) \;. \ee
Since $\Phi(-\nu_k^2)$ is Hermitian, we can apply standard
perturbation techniques to our problem. The eigenvalue problem for
the principal matrix we wish to solve is given in (\ref{eigenvalue
principal delta}).  We again suppose that there is no degeneracy
for simplicity. The energy eigenvalue changes to $E_k = E_k^{(0)}
+ \delta E_k$ or
\be \nu_k = \nu_k^{(0)} + \delta \nu_k \;. \ee
From the fundamental idea of the perturbation theory in finite
dimensional spaces, we have the following expansions for the
eigenvalues and eigenvectors:
\beqs \omega^{k} &=& \omega^{k(0)} + \omega^{k(1)} + \omega^{k(2)}
+ \ldots \;,  \cr A^{k}_{i} &=& A^{k(0)}_{i} + A^{k(1)}_{i} +
A^{k(2)}_{i} + \ldots \;, \eeqs
and the solution to the related unperturbed eigenvalue problem
\be \sum_{j=1}^{N} \left[\Phi_{D}(-\nu_k^2)\right]_{ij}
A_{j}^{k(0)}(-\nu_k^2) = \omega^{k(0)}(-\nu_k^2)
A_{i}^{k(0)}(-\nu_k^2) \;, \ee
is given by
\be \omega^{k(0)}(-\nu_k^2) = \int_0 ^\infty {\mathrm{d} t \over
\hbar} K_{t}(a_k,a_k;g)[e^{-t\mu_k^2/\hbar} -e^{-t \nu_k^2/\hbar}]
\;. \ee
Then, the energy eigenvalues can easily be found from the
condition $\omega^{k(0)}(-\nu_k^2)=0$:
\be E_k^{(0)}=-\mu_k^2 \hspace{1cm} \mathrm{or} \hspace{1cm}
\nu_k^{(0)} = \mu_k \;, \ee
and eigenvectors are
\begin{equation} A^{k(0)}(\mu_k) \equiv A^{k(0)} \equiv
\mathbf{e}^{k} \equiv
\left(%
\begin{array}{cc}
  0 \cr \vdots \cr 1 \cr \vdots \cr 0
\end{array}%
\right)\;,
\end{equation}
where $1$ is located in the $k$th position of the column and other
elements of it are zero or we can write $A^{k(0)}_i = e^{k}_i=
\delta_{k i}$. Here $e^{k}_{i}$ s form a complete orthonormal set
of basis.
\be \sum_{i=1}^{N} e^{k}_{i}e^{l}_{i}= \delta_{kl} \;. \ee
We must emphasize that there is no distinction between upper and
lower indices for our purposes. Once we have found the solution of
the diagonal part of the principal matrix or unperturbed
eigenvalue problem, we can perturbatively solve the whole problem.
The standard perturbation theory gives us the first and second
order eigenvalues:
\beqs \omega^{k(1)} (-\nu_k^2) &=& \sum_{i,j=1}^{N} e^{k}_{i}
\left[\delta \Phi(-\nu_k^2)\right]_{ij} e^{k}_{j} = \left[\delta
\Phi(-\nu_k^2)\right]_{kk} =0 \;, \cr \omega^{k(2)}(-\nu_k^2) &=&
\sum_{\substack{l=1 \\
l \neq k}}^N { \left| \sum_{i,j=1}^{N} e^{l}_{i} \left[\delta
\Phi(-\nu_k^2)\right]_{ij} e^{k}_{j}\right|^2 \over
\omega^{k(0)}(-\nu_k^2)-\omega^{l(0)}(-\nu_k^2)} \cr &=& \sum_{\substack{l=1 \\
l \neq k}}^N {\Phi_{lk}(-\nu_k^2) \Phi_{k l}(-\nu_k^2) \over
\omega^{k(0)}(-\nu_k^2)-\omega^{l(0)}(-\nu_k^2) },
\label{perturbation eigenvalues} \eeqs
respectively. Hence the energy eigenvalues of the whole problem
can be determined from
\be \omega^{k}(-\mu_k^2 + \delta E_k) = \omega^{k(0)}(-\mu_k^2 +
\delta E_k) + \omega^{k(2)}(-\mu_k^2 + \delta E_k) + \ldots = 0
\;, \label{perturb eigenvalue eq}\ee
and $\omega^{k(0)}(-\nu_k^2)$ and $\Phi_{kl}(-\nu_k^2)$ for $k
\neq l$ can be expanded around $\nu_k=\mu_k$
\beqs  \omega^{k(0)}(-\mu_k^2 + \delta E_k) &=&  \left. {\partial
\omega^{k(0)}(-\nu_k^2) \over
\partial \nu_k} \right|_{\nu_k=\mu_k} \delta \nu_k + \mathcal{O}(\delta^2 \nu_k) \;, \cr
\Phi_{kl}(-\mu_k^2 + \delta E_k) &=& \Phi_{kl}(-\mu_k^2) + \left.
{\partial \Phi_{kl}(-\nu_k^2) \over \partial \nu_k}
\right|_{\nu_k=\mu_k} \delta \nu_k + \mathcal{O}(\delta^2 \nu_k)
\;, \label{perturb taylor exp} \eeqs
where we have used the fact $\omega^{k(0)}(-\mu_k^2)=0$. If we
substitute (\ref{perturb taylor exp}) into (\ref{perturb
eigenvalue eq}) and (\ref{perturbation eigenvalues}), and use
Feynman-Hellman theorem (\ref{feynmanhelmann}), we obtain
\beqs & & \hskip-1cm 0 = \left. {\partial \Phi_{kk}(-\nu_k^2)
\over
\partial
\nu_k} \right|_{\nu_k=\mu_k} \delta \nu_k - \sum_{\substack{l=1 \\
l \neq k}}^N {1 \over \Phi_{ll}(-\mu_k^2)}
\Bigg[\Phi_{kl}(-\mu_k^2) \Phi_{lk}(-\mu_k^2) \cr & & + \Bigg(
\Phi_{kl}(-\mu_k^2) \left. {\partial \Phi_{lk}(-\nu_k^2) \over
\partial \nu_k} \right|_{\nu_k=\mu_k} + \Phi_{lk}(-\mu_k^2) \left. {\partial \Phi_{kl}(-\nu_k^2) \over
\partial \nu_k} \right|_{\nu_k=\mu_k}\Bigg) \delta \nu_k \Bigg] \cr & & \hskip-1cm \times \Bigg[1 +
{1 \over \Phi_{ll}(-\mu_k^2)} \Bigg( \left. {\partial
\Phi_{ll}(-\nu_k^2) \over
\partial \nu_k} \right|_{\nu_k=\mu_k}- \left. {\partial \Phi_{kk}(-\nu_k^2) \over
\partial \nu_k} \right|_{\nu_k=\mu_k} \Bigg) \delta \nu_k \Bigg]^{-1}  + \mathcal{O}(\delta^2 \nu_k) \;. \eeqs
If we also expand the last factor in the powers of $\delta \nu_k$
and ignore the second order terms and combine the terms using the
symmetry property of principal matrix, we find
\beqs & & \Bigg[ \left. {\partial \Phi_{kk}(-\nu_k^2) \over
\partial
\nu_k} \right|_{\nu_k=\mu_k} + \sum_{\substack{l=1 \\
l \neq k}}^N  {\Phi_{kl}(-\mu_k^2) \Phi_{lk}(-\mu_k^2) \over
\Phi_{ll}^2(-\mu_k^2)} \Bigg( \left. {\partial \Phi_{ll}(-\nu_k^2)
\over
\partial \nu_k} \right|_{\nu_k=\mu_k} \cr & & \hspace{2cm} - \left. {\partial \Phi_{kk}(-\nu_k^2) \over
\partial \nu_k} \right|_{\nu_k=\mu_k} \Bigg) - 2 \sum_{\substack{l=1 \\
l \neq k}}^N  {\Phi_{kl}(-\mu_k^2) \over \Phi_{ll}(-\mu_k^2)}
\left. {\partial \Phi_{lk}(-\nu_k^2) \over
\partial
\nu_k} \right|_{\nu_k=\mu_k} \Bigg] \delta \nu_k \cr & =& \sum_{\substack{l=1 \\
l \neq k}}^N  {\Phi_{kl}(-\mu_k^2) \Phi_{lk}(-\mu_k^2) \over
\Phi_{ll}(-\mu_k^2)} + \mathcal{O}(\delta^2 \nu_k)\;. \eeqs
Ignoring the second and third terms on the left hand side of the
equality due to the fact that $\Phi_{kk}(-\nu_k^2) \gg
|\Phi_{kl}(-\nu_k^2)|$, we get the change in $\nu_k$
\be \delta \nu_k \approx   \left(\left. {\partial
\Phi_{kk}(-\nu_k^2) \over
\partial
\nu_k} \right|_{\nu_k=\mu_k} \right)^{-1} \sum_{\substack{l=1 \\
l \neq k}}^N  {\Phi_{kl}(-\mu_k^2) \Phi_{lk}(-\mu_k^2) \over
\Phi_{ll}(-\mu_k^2)} + \mathcal{O}(\delta^2 \nu_k)\;,
\label{perturbation result} \ee
so the change in the energy is $\delta E_k \approx -2\mu_k \delta
\nu_k + \mathcal{O}(\delta^2 \nu_k)$.
Let us now consider how the bound state energy changes in the
tunneling regime for our problem in which ${\hbar^2 \over 2m
d_{ij}^{2}} \ll \mu_{k,min}^{2}$. We must first calculate
asymptotic behavior of the off-diagonal element of the principal
matrix in this regime. In order to see this, we make the scaling
transformation $t=u/B$, where $B= \hbar/ 2m d_{ij}^{2}$ and use
the scaling property of the heat kernel
(\ref{heatkernelscaling2}). In two dimensions, we have
\beqs \Phi_{ij}(-\mu_k^2) = - \int_{0}^{\infty} {\mathrm{d} u
\over \hbar} \; K_{u}(a_i,a_j;B g) e^{- {u \mu_k^2 \over \hbar B}}
\;. \eeqs
In tunnelling regime, the most significant contribution to the
integral comes from the region $u=0$ due to the fact that
integrand is suppressed by the exponential term for large values
of $u$. Hence we can use the short time asymptotic of the heat
kernel given in (\ref{molchanov}). The result is an integral
representation of the modified Bessel function of the third kind
\cite{lebedev}
\beqs \Phi_{ij}(-\mu_k^2) \sim -  {2 d_{ij}^{1/2}  \over (4 \pi
\hbar^2 /2 m)} K_0\left(\sqrt{{2 m d_{ij}^2 \mu_k^2 \over
\hbar^2}}\right) \sum_l \Psi_{l}^{-1/2}(x,y) \;, \eeqs
where we have used $d_{ij} \rightarrow B^{1/2} d_{ij}$ and $\Psi_l
\rightarrow \Psi_l / B^{1/4}$ (for two dimensions) under the
scaling transformation $g\rightarrow B g$. The asymptotic
expansion of $K_0(x)$ for large values of $x$
\be K_0(x) \sim \sqrt{\pi \over 2 x} e^{-x} \ee
leads to
\beqs \Phi_{ij}(-\mu_k^2) \sim - \sqrt{{\pi \over 2}} { \sum_l
\Psi_{l}^{-1/2}(x,y)  \over (\pi \hbar^2 / m)}  \left({\hbar^2
\over 2 m \mu_k^2} \right)^{1/4} e^{-{\sqrt{2m}d_{ij} \mu_k \over
\hbar}} \label{asymp Phi 2d} \;. \eeqs
Here, large values of $x$ corresponds to tunneling regime in our
problem. For three dimensional case, the idea is the same and the
result would be
\beqs \Phi_{ij}(-\mu_k^2) \sim - \sqrt{{2 \pi \over m \hbar }} {
\sum_l \Psi_{l}^{-1/2}(x,y)  \over (4 \pi \hbar / 2m)^{3/2}}
e^{-{\sqrt{2m}d_{ij} \mu_k \over \hbar}} \label{asymp Phi 3d} \;.
\eeqs
Therefore, in the tunnelling regime, we can find the change in the
bound state energy $\delta E_k$ in the presence of other delta
interactions by substituting (\ref{asymp Phi 2d}) and (\ref{asymp
Phi 3d}) into (\ref{perturbation result}). In agreement with the
naive expectation in the standard quantum mechanics, we show that
the bound state energy in the tunnelling regime gets exponentially
smaller with increasing distance between the centers.

Before we prove the lower bound for the ground state energy and
find pointwise bounds on the wave functions, the results in
mathematics literature for the upper and lower bounds of the heat
kernel  for compact and Cartan-Hadamard manifolds will be briefly
given in the next section.

\section{Upper and Lower Bound Estimates of the Heat Kernel}
\label{Upper and Lower Bound Estimates of the Heat Kernel}

\subsection{Compact Manifolds:}
\label{heat kernel compact}

1) Upper Bound of Diagonal Heat Kernel: The following result is a
simplified version of the corollary 3.6 given in \cite{wang},
which assumes that some geometrical conditions must hold on the
boundary. The global upper bound estimate of the diagonal heat
kernel in \cite{wang} includes whole boundary information via an
explicitly calculable strictly positive constant $A \equiv
A(\mathrm{diam}(\mathcal{M}), H, R, K,V({\mathcal M}))$, where
$\mathrm{diam}(\mathcal{M})$ is the diameter of the manifold, and
$K$ is the lower bound on the Ricci curvature, and also $H$ and
$R$ are parameters related to boundary conditions. We then state
the following corollary by safely removing the boundary effects in
corollary 3.6 in \cite{wang} since $A$ is strictly positive:

Let $\mathcal M$ be a compact manifold. Suppose that the Ricci
curvature of $\mathcal M$ satisfies $\mathrm{Ric}_{\mathcal M}
\geq -K, K\geq 0$. Then $\forall t>0$ and $x \in \mathcal M$
\be \label{compactdiagupp} K_t(x,x;g) \leq \frac{1}{V(\mathcal
M)}+A' (\hbar t/2m)^{-D/2} \;, \ee
where $A' \equiv A'(\mathrm{diam}(\mathcal{M}), K, V({\mathcal
M}))$.

2) Upper Bound of Off-Diagonal Heat Kernel:  The following
corollary given in \cite{grigoryan2} constrains the off-diagonal
elements of the heat kernel from above.

Assume that for some points $x,y \in {\mathcal M}$ ($\mathcal{M}$
is any Riemannian manifold) and $\forall t>0,$
\be \label{func} K_t(x,x;g) \leq \frac{C}{f(t)} \mbox{   and   }
K_t(y,y;g) \leq \frac{C}{g(t)} \;, \ee
where $f$ and $g$ are increasing positive functions on $(0,\infty
)$ satisfying the regularity condition given below. Then, for any
$C_2 >2$ and for all $t>0$
\be \label{offdiagupp} K_t(x,y;g) \leq
\frac{4A}{\sqrt{f(\varepsilon \, t) g(\varepsilon \, t)}} \
\mbox{exp} \left(-\frac{m d^2(x,y)}{\hbar C_2 t}\right)\;, \ee
where $\varepsilon =\varepsilon(C_2,a)$, $A$ and $a$ are the
constants from the regularity condition below.

\textit{Regularity Condition:} There are numbers $A \geq 1$ and
$a>1$ such that
\be \label{regularity} \frac{f(a s)}{f(s)} \leq A \frac{f(a
t)}{f(t)} \;,\ee
for all $0<s<t$.

By comparing the equations (\ref{compactdiagupp}) and
(\ref{func}), we realized that the right hand side of
(\ref{compactdiagupp}) can be an explicit candidate for the
functions $f$ or $g$ in the theorem above. Hence, we could have
\be \label{regularfunc} f(t)=g(t)=\left[\frac{1}{V(\mathcal M)}+A'
(\hbar t/2m)^{-D/2}\right]^{-1} \;, \ee
by choosing $C=1$. It is easy to check that these functions are
positive and increasing. We can also verify that they satisfy the
regularity condition (\ref{regularity}) with $A=a^{D/2}$.
Therefore, we have obtained the upper bound for the off-diagonal
elements of the heat kernel
\begin{eqnarray}
\label{compactoffdiagupp} K_t(x,y;g) \leq 4A
\left[\frac{1}{V(\mathcal M)} + B(\varepsilon) (\hbar
t/2m)^{-D/2}\right] \mbox{exp} \left(-\frac{m d^2(x,y)}{\hbar C_2
t}\right)\;,
\end{eqnarray}
where $B(\varepsilon)=A' \varepsilon^{-D/2}$.

3) The Lower Bound of Heat Kernel:

We have a direct theorem about the lower bound on the heat kernel
(cf. theorem 5.6.1 in \cite{Davies}). We will just give the
statement of the theorem: Let ${\mathcal M}$ be a complete
Riemannian manifold with $\mathrm{Ric}_{\mathcal{M}} \geq 0$.
Then, we have
\be \label{compactlow} K_t(x,y;g) \geq (4 \pi \hbar t/2m)^{-D/2}
\exp \left( - {m d^2(x,y) \over 2 \hbar t} \right) \;,\ee
for all $x,y \in {\mathcal M}$ and $t>0$. In particular, we find
the lower bound to be
\be \label{compactdiaglow} K_t(x,x;g) \geq (4 \pi \hbar
t/2m)^{-D/2}\;. \ee

\subsection{Cartan-Hadamard Manifolds:}
\label{heat kernel noncompact}

A manifold ${\mathcal M}$ is called a Cartan-Hadamard manifold
\cite{grigoryan} if ${\mathcal M}$ is a geodesically complete,
simply connected, non-compact Riemannian manifold with
non-positive sectional curvature everywhere. The $D$-dimensional
flat ${\mathbb R}^D$ and hyperbolic spaces ${\mathbb H}^D$ are the
best known examples of Cartan-Hadamard manifolds.

1) The Upper Bound of Heat Kernel: In order to give an upper bound
for Cartan-Hadamard manifolds, we need to give the some
definitions and related theorems in the literature.

\textit{Isoperimetric Inequalities:} Isoperimetric inequalities
are the relations between the boundary area of regions and their
volume. We say that manifold $\mathcal{M}$ admits the
isoperimetric function $I$ if for any precompact open set $\Omega
\subset \mathcal{M}$ with smooth boundary
\be A(\partial \Omega) \geq I(v) \;, \ee
where $A(\partial \Omega)$ is the area of boundary of the region
$\Omega$, and $v=V(\Omega)$ is the volume of the region. Any
Cartan-Hadamard manifold ${\mathcal M}$ of dimension $D$ admits
the isoperimetric function $I(v)=\kappa v^{\frac{D-1}{D}}$,
$\kappa>0$ \cite{hoffman}. We have an important theorem
\cite{grigoryan3} given below:

Assume that manifold ${\mathcal M}$ admits a non-negative
continuous isoperimetric function $I(v)$ such that $I(v)/v$ is
non-increasing. Let us define the function $f(t)$ by
\be \label{function} t = 4 \int_0^{f(t)} \mathrm{d} v \frac{v
}{I^2(v)} \;, \ee
assuming that the integral does not divergent at $t=0$. Then for
all $x \in \mathcal{M}$, $t>0$ and $\varepsilon >0$,
\be K_t(x,x;g) \leq {2 \varepsilon^{-1} \over f((1-\varepsilon)t)}
\;. \ee
Moreover, if the function $f$ satisfies in addition the regularity
condition (\ref{regularity}) then, for all $x,y\in {\mathcal
M},t>0 ,C_2>2$ and some $\varepsilon >0$,
\be K_t(x,y;g) \leq \frac{4 A}{f(\varepsilon t)} \mbox{ exp}
\left(\frac{-m d^2(x,y)}{\hbar C_2 t}\right) \;.\ee
The isoperimetric function for Cartan-Hadamard manifolds given
above satisfies all the requirements above, that is, it is
non-negative continuous function and $I(v)/v$ is non-increasing.
Substituting the above isoperimetric function for Cartan-Hadamard
manifolds into (\ref{function}) we obtain the function $f(t)$ by a
simple integration
\be f(t)=\left(\frac{\kappa^2}{2D} t \right)^{D/2} \;, \ee
and it meets all the requirements of this theorem including the
regularity condition. Hence, the upper bound on the off-diagonal
elements of the heat kernel on Cartan-Hadamard manifolds is given
by
\be \label{hadamardoffdiagupp} K_t(x,y;g) \leq \frac{C(\varepsilon
,\kappa)}{(4\pi \hbar t/2m)^{D/2}} \mbox{ exp} \left(\frac{-m
d^2(x,y)}{\hbar C_2 t}\right) \;, \ee
where $C(\varepsilon,\kappa)= {4 A \over (\kappa^2
\varepsilon/2D)^{D/2}}$ and the physical parameters $\hbar,m$ are
introduced for dimensional reasons. This upper bound estimate of
the heat kernel is also valid for minimal submanifolds, which are
submanifolds of $\mathbb{R}^D$ whose normal mean curvature vector
$ H(x)=(H_1(x),H_2(x),...,H_D(x))$ vanishes for all $x \in
\mathcal{M}$, because minimal submanifolds admit the same form of
isoperimetric function \cite{bombieri} with the Cartan-Hadamard
manifolds.

2) The Lower Bound of the Diagonal Heat Kernel: On Cartan-Hadamard
manifolds, the lower bound of the diagonal elements of the heat
kernel are obtained in \cite{grigoryan}. The lower bound of the
diagonal heat kernel is given also for any manifold given as a
proposition (5.14) in \cite{grigoryan}: For any manifold
$\mathcal{M}$, for any $x \in \mathcal{M}$ and $\delta > 0$, there
exists $c=c_x >0$ such that
%
%Assume that the sectional curvature inside some ball $B(x,r)$ is
%bounded below by $-K^2_{max}(r)$, where $x \in \mathcal{M}$. Then,
%for all $t>0$ and $\delta>0$,
%
\be \label{hadamarddiaglow} K_t(x,x;g) \geq \frac{c}{(4\pi \hbar
t/2m)^{D/2}} \mbox{exp} \left[-(\sigma_1({\mathcal
M})+\delta){\hbar t \over 2m} \right] \;, \ee
where $\sigma_1({\mathcal M})$ is the spectral radius
corresponding to the Laplacian and it is restricted to the
following range \cite{docarmo, chavel2}
\be \frac{1}{4} (D-1)^2 K_{max}^2 \geq \sigma_1({\mathcal M}) \geq
\frac{1}{4} (D-1)^2 K_{min}^2 \;, \ee
for a Cartan-Hadamard manifold whose sectional curvature is
bounded from above by $-K_{min}^2$.

\section{Pointwise Bounds on Wave function}
\label{Pointwise Bounds on Wave function}

There is extensive amount of literature on the exponential decays
of the wave functions of the Schr\"{o}dinger operators, which
states that $L^2$ solutions of $(- \nabla^2 + V)\psi=E \psi $ obey
pointwise bounds of the form
\be |\psi(r)| \leq C_a e^{-a r} \;, \ee
if the potential energy $V$ is continuous and bounded below and
$E$ is in the discrete spectrum of $- \nabla^2 + V$ (see
\cite{simon} for the review of the subject). The proofs given in
the literature do not include the potentials which require
renormalization and they are valid only for $\mathbb{R}^D$. We
shall prove that it is still possible to get exponential pointwise
bounds for our problem.

It is easy to see the upper bound of the wave function
(\ref{wavefunction heat delta}) by applying Cauchy Schwartz
inequality
\beqs |\psi_k(x)| & \leq &  \alpha  \left| \sum_{i=1}^N
A_i(-\nu_k^2) \int_{0}^{\infty} {\mathrm{d} t \over \hbar}\;
e^{-\frac{t \nu_{k}^2}{\hbar}} K_{t}(a_i, x;g) \right| \cr & \leq
& \alpha  \left[ \sum_{i=1}^N \left|\int_{0}^{\infty} {\mathrm{d}
t \over \hbar}\; e^{-\frac{t \nu_{k}^2}{\hbar}} K_{t}(a_i,
x;g)\right|^2 \right]^{1/2} \cr & \leq & \alpha \sum_{i=1}^N
\int_{0}^{\infty} {\mathrm{d} t \over \hbar}\; e^{-\frac{t
\nu_{k}^2}{\hbar}} K_{t}(a_i, x;g) \;, \eeqs
where we call the coefficient (which is independent of the
coordinates $x$) in front of the integral in (\ref{wavefunction
heat delta}) as $\alpha$ and we have used the fact that
$\sum_{i=1}^{N}|A_i(-\nu_k^2)|^2 = 1$. Thanks to the upper bound
on the heat kernel given in (\ref{compactoffdiagupp}) and
(\ref{hadamardoffdiagupp}), we show that the wave function is
pointwise bounded on $\mathcal{M}$. For compact manifolds, the
upper bound (\ref{compactoffdiagupp}) of heat kernel gives
\beqs |\psi_k(x)| & \leq &  8 \alpha A \sum_{i=1}^N \Bigg[ {1
\over V(\mathcal{M})} \sqrt{{m d^2(a_i,x) \over \nu_k^2 \hbar^2
C_2 }} K_1 \left( 2 \sqrt{{m d^2(a_i,x) \nu_k^2 \over  \hbar^2 C_2
}} \right) \cr & & \hskip-1cm + {B(\varepsilon) \over \hbar
\left(\hbar/2m \right)^{D/2}} \left({m d^2(a_i,x) \over \nu_k^2
C_2}\right)^{{1 \over 2}-{D \over 4}} K_{{D \over 2}-1} \left(2
\sqrt{{m d^2(a_i,x) \nu_k^2 \over \hbar^2 C_2 }} \right) \Bigg]
 \;, \label{psiboundcompact1} \eeqs
where we use the following integral representation of the modified
Bessel functions of the third kind (or sometimes called
Macdonald's functions) \cite{lebedev}
\be K_{\nu}(z)= {1 \over 2} \Bigg({z \over 2}\Bigg)^{\nu}
\int_{0}^{\infty} \mathrm{d} s \; e^{-s-(z^2/4s)} s^{-\nu-1} \;
\hspace{1cm} |\mathrm{arg}z|< {\pi \over 4}, \; \mathrm{Re} (\nu)
> -{1 \over 2} \;. \label{intrepbessel 1} \ee
For $D=2$ and $D=3$, we can also find the upper bounds on Bessel
functions $K_0$ and $K_1$ with another useful integral
representation \cite{lebedev}:
\be K_{\nu}(z) = {\sqrt{\pi} z^{\nu} \over 2^{\nu}
\Gamma(\nu+{1\over 2})} \int_{0}^{\infty} \mathrm{d} s \; e^{-z
\cosh s} \sinh^{2 \nu} s \;, \hspace{1cm} \mathrm{Re}(z)> 0, \;
\mathrm{Re} (\nu) > -{1 \over 2} \;. \label{intrepbessel 2} \ee
Using the inequality $\cosh s = {e^s + e^{-s} \over 2} > {e^{s}
\over 2}$ for all $s \geq 0$ in (\ref{intrepbessel 2}), we have an
upper bound $K_0(x)$ for $x \in \mathbb{R}^{+}$
\beqs K_{0}(x) & < & \int_{0}^{\infty} \mathrm{d} s \; e^{-{x
\over 2} e^{s} } \;. \eeqs
By subsequent change of variables $\xi = e^s $ and $\eta = \xi - 1
$, we get
\beqs K_{0}(x) & < & \int_{0}^{\infty} \mathrm{d} \eta \; { e^{-{x
(\eta +1 ) \over 2}} \over \eta + 1}  \;. \eeqs
If we also define a new variable $z= x \eta$, we have
\beqs K_{0}(x) & < & e^{-{x \over 2}} \int_{0}^{\infty} \mathrm{d}
z \; { e^{-{z \over 2}} \over z + x }  \leq   {e^{-{x \over 2}}
\over x} \int_{0}^{\infty} \mathrm{d} z \;  e^{-{z \over 2}} \cr &
\leq & {2 \over x} \; e^{-{x \over 2}} \label{bessel0estiamte} \;.
\eeqs
Alternatively, we can find a sharper bound for $K_0$ if we divide
the integral above into two parts as follows:
\beqs K_{0}(x) & < &  e^{-{x \over 2}} \left[ \int_{0}^{1}
\mathrm{d} z \;  {e^{-{z \over 2}} \over z+x} +  \int_{1}^{\infty}
\mathrm{d} z \;  {e^{-{z \over 2}} \over z+x} \right] \cr & \leq &
e^{-{x \over 2}} \left[ \int_{0}^{1} \mathrm{d} z \;  {1 \over
z+x} +  {1 \over 1+ x} \int_{1}^{\infty} \mathrm{d} z \;  e^{-{z
\over 2}} \right] \;. \eeqs
Hence, we have the following bound for $K_0$ which shows the
logarithmic singularity near $x=0$
\beqs K_0(x) & \leq  & e^{-{x \over 2}} \left[ \ln \left({x+1
\over x}\right) + {2(1-e^{-1/2}) \over 1+ x} \right] \cr  & \leq &
{2 \over 1+ x} e^{-{x \over 2}} + e^{-{x \over 2}} \ln \left({x+1
\over x}\right)\;. \label{bessel0sharperestimate} \eeqs
Using $\sinh^2 s = ({e^s -e^{-s} \over 2})^2 < {e^{2s} \over 4} $
for all $s\geq 0$ and following the same steps above for $K_1$, we
find
\be \label{bessel1estimates} K_{1}(x) \leq e^{-\frac{x}{2}}
\left(\frac{1}{x}+\frac{1}{2}\right)\;. \ee
Substituting (\ref{bessel0sharperestimate}) and
(\ref{bessel1estimates}) into (\ref{psiboundcompact1}) for $D=2$,
we get
\newpage
\beqs & & \hskip-1cm |\psi_k(x)| \leq 8 \alpha A \sum_{i=1}^N
\Bigg[ {1 \over 2 V(\mathcal{M})} \sqrt{{m d^2(a_i,x) \over
\nu_k^2 \hbar^2 C_2 }} e^{-\sqrt{m d^2(a_i,x) \nu_k^2 / \hbar^2
C_2}} \left({1 \over \sqrt{m d^2(a_i,x) \nu_k^2 / \hbar^2 C_2}}+1
\right) \cr & & + {B(\varepsilon) \over \hbar \left(\hbar/2m
\right)} e^{-\sqrt{m d^2(a_i,x) \nu_k^2 / \hbar^2 C_2}}\Bigg(\ln
\left( { 2 \sqrt{m d^2(a_i,x) \nu_k^2 / \hbar^2 C_2} + 1 \over 2
\sqrt{m d^2(a_i,x) \nu_k^2 / \hbar^2 C_2}} \right) \cr & & + {2
\over 1+ 2 \sqrt{m d^2(a_i,x) \nu_k^2 / \hbar^2 C_2}} \Bigg)
\Bigg] \;, \label{psiboundcompactD2} \eeqs
and for $D=3$, we have
\beqs |\psi_k(x)| & \leq &  8 \alpha A \sum_{i=1}^N \Bigg[ {1
\over 2 V(\mathcal{M})} \sqrt{{m d^2(a_i,x) \over \nu_k^2 \hbar^2
C_2 }} e^{-\sqrt{m d^2(a_i,x) \nu_k^2 / \hbar^2 C_2}} \cr & &
\hskip-2cm \times \left({1 \over \sqrt{m d^2(a_i,x) \nu_k^2 /
\hbar^2 C_2}}+1 \right) + {B(\varepsilon) \sqrt{\pi} \over \hbar
\left(\hbar/2m \right)^{3/2}}
  {e^{-2 \sqrt{m
d^2(a_i,x) \nu_k^2 / \hbar^2 C_2}} \over 2 \sqrt{m d^2(a_i,x)
/\hbar C_2}} \Bigg]\;, \label{psiboundcompactD3} \eeqs
where we have used the explicit exact expression for $K_{{1\over
2}}$
\be K_{{1 \over 2}}(u)= \sqrt{{\pi \over 2 u}} e^{-u} \;. \ee
We can repeat the same steps for Cartan-Hadamard manifolds by
using the upper bounds of heat kernel given in
(\ref{hadamardoffdiagupp}) and the result is
\beqs |\psi_k(x)| & \leq & 2 \alpha \sum_{i=1}^N \Bigg[
{C(\varepsilon, \kappa) \over \hbar (4 \pi \hbar /2m)}  e^{-
\sqrt{m d^2(a_i,x) \nu_k^2 / \hbar^2 C_2}}  \Bigg(\ln \left( { 2
\sqrt{m d^2(a_i,x) \nu_k^2 / \hbar^2 C_2} + 1 \over 2 \sqrt{m
d^2(a_i,x) \nu_k^2 / \hbar^2 C_2}} \right) \cr & &
 + {2 \over 1+ 2 \sqrt{m d^2(a_i,x) \nu_k^2 / \hbar^2 C_2}}
\Bigg) \Bigg] \;, \label{psiboundcartanhadamardD2}\eeqs
for $D=2$ and
\be |\psi_k(x)| \leq  \alpha \sum_{i=1}^N \Bigg[ {C(\varepsilon,
\kappa) \sqrt{\pi}\over \hbar (4 \pi \hbar /2m)^{3/2}} { e^{-2
\sqrt{m d^2(a_i,x) \nu_k^2 / \hbar^2 C_2}} \over \left(m
d^2(a_i,x) \nu_k^2 / \hbar^2 C_2 \right)^{1/4}} \Bigg] \;,
\label{psiboundcartanhadamardD3}\ee
for $D=3$. In standard quantum mechanics, the pointwise
exponential bounds do not take into account singular interactions.
Nevertheless, we prove that they are still valid for our problem
in two and three dimensions.

In order to understand heuristically why our problem can be
considered as a self-adjoint extension, which is also suggested by
Krein's formula obtained in section \ref{An Alternative
Construction of the Model}, it is interesting to calculate the
expectation value of the free energy for the bound state. The
result (\ref{wavefunction heat delta}) permits us to write the
expectation value as
\beqs & & \langle \psi_k | H_0 | \psi_k \rangle = \left[
\sum_{i,j=1}^N A_i^{*}(-\nu_k^2) \int_{0}^{\infty} {\mathrm{d} t
\over \hbar} \; \left({t \over \hbar} \right) K_t(a_i,a_j;g)
e^{-\frac{t \nu_{k}^2}{\hbar}} \, A_j(-\nu_k^2) \right]^{-1} \cr &
& \int_{\mathcal{M}} \mathrm{d}_{g}^{D} x \; \Bigg[
\int_{0}^{\infty} {\mathrm{d} t_1 \over \hbar} \; e^{- {t_1
\nu_k^2 \over \hbar}} \sum_{i=1}^{N} A_{i}^{*}(-\nu_k^2)
K_{t_1}(a_i,x;g) \Bigg] \Bigg[ \int_{0}^{\infty} {\mathrm{d} t_2
\over \hbar} \; e^{- {t_2 \nu_k^2 \over \hbar}} \sum_{j=1}^{N}
A_j(-\nu_{k}^{2}) \cr & & \left(-\hbarm \nabla_{g}^{2}
K_{t_2}(a_j,x;g) \right)\Bigg] \;.\eeqs
Using (\ref{heatkernelequation}) and making an integration by
parts to $t_2$ integral, we have
\beqs & & \langle \psi_k | H_0 | \psi_k \rangle = - \left[
\sum_{i,j=1}^N A_i^{*}(-\nu_k^2) \int_{0}^{\infty} {\mathrm{d} t
\over \hbar} \; \left({t \over \hbar} \right) K_t(a_i,a_j;g)
e^{-\frac{t \nu_{k}^2}{\hbar}} \, A_j(-\nu_k^2) \right]^{-1} \cr &
& \int_{\mathcal{M}} \mathrm{d}_{g}^{D} x \; \Bigg[
\int_{0}^{\infty} {\mathrm{d} t_1 \over \hbar} \; e^{- {t_1
\nu_k^2 \over \hbar}} \sum_{i=1}^{N} A_i^{*}(-\nu_k^2)
K_{t_1}(a_i,x;g) \Bigg] \Bigg[ - \sum_{j=1}^{N} A_j(-\nu_k^2)
\delta_g(x,a_j) \cr & & +  \sum_{j=1}^{N} A_j(-\nu_k^2)
\int_{0}^{\infty} {\mathrm{d} t_2 \over \hbar} \; \nu_k^2 \; e^{-
{t_2 \nu_k^2 \over \hbar}} K_{t_2}(a_j,x;g) \Bigg] \;,\eeqs
where we have used the initial condition of heat kernel
(\ref{initialcon}). Integrating with respect to $x$ and using the
semigroup property of heat kernel (\ref{semigroupprop}), we obtain
\beqs & & \langle \psi_k | H_0 | \psi_k \rangle = \left[
\sum_{i,j=1}^N A_i^{*}(-\nu_k^2) \int_{0}^{\infty} {\mathrm{d} t
\over \hbar} \; \left({t \over \hbar} \right) K_t(a_i,a_j;g)
e^{-\frac{t \nu_{k}^2}{\hbar}} \, A_j(-\nu_k^2) \right]^{-1} \cr &
&  \Bigg[ \int_{0}^{\infty} {\mathrm{d} t_1 \over \hbar} \; e^{-
{t_1 \nu_k^2 \over \hbar}} \sum_{i,j=1}^{N} A_i^{*}(-\nu_k^2) \;
A_j(-\nu_k^2) K_{t_1}(a_i,a_j;g) \cr & & - \int_{0}^{\infty}
{\mathrm{d} t_1 \over \hbar} \int_{0}^{\infty} {\mathrm{d} t_2
\over \hbar} \; \sum_{i,j=1}^{N} A_i^{*}(-\nu_k^2) A_j(-\nu_k^2)
K_{t_1+t_2}(a_i,a_j;g) e^{-{(t_1+t_2)\nu_k^2 \over \hbar}} \nu_k^2
\Bigg] \;.\eeqs
By change of variables $u=t_1+t_2$ and $v=t_1-t_2$, we find
\beqs & & \langle \psi_k | H_0 | \psi_k \rangle = \left[
\sum_{i,j=1}^N A_i^{*}(-\nu_k^2) \int_{0}^{\infty} {\mathrm{d} t
\over \hbar} \; \left({t \over \hbar} \right) K_t(a_i,a_j;g)
e^{-\frac{t \nu_{k}^2}{\hbar}} \, A_j(-\nu_k^2) \right]^{-1} \cr &
&  \Bigg[ \int_{0}^{\infty} {\mathrm{d} t_1 \over \hbar} \; e^{-
{t_1 \nu_k^2 \over \hbar}} \sum_{i,j=1}^{N} A_i^{*}(-\nu_k^2) \;
A_j(-\nu_k^2) K_{t_1}(a_i,a_j;g) \cr & & - {1 \over 2}
\int_{0}^{\infty} {\mathrm{d} u \over \hbar^2} \;
\Bigg(\int_{-u}^{u} \mathrm{d} v \Bigg) \; \sum_{i,j=1}^{N}
A_i^{*}(-\nu_k^2) A_j(-\nu_k^2) K_{u}(a_i,a_j;g) e^{-{u \nu_k^2
\over \hbar}} \nu_k^2 \Bigg] \;.\eeqs
One can easily see that the $i=j$ term of the sum for the first
term in the parenthesis
\be \int_{0}^{\infty} {\mathrm{d} t_1 \over \hbar} \; e^{- {t_1
\nu_k^2 \over \hbar}} |A_i(-\nu_k^2)|^2 K_{t_1}(a_i,a_i;g) \ee
is divergent due to diagonal short time asymptotics of heat kernel
(\ref{asymheat}) for $D\geq 2$ and one can also show that the off-
diagonal terms of this sum in the parenthesis and all terms in
second sum is convergent by upper bound on the heat kernel given
in (\ref{compactoffdiagupp}) and (\ref{hadamardoffdiagupp}). Hence
we find that the expectation value of the free Hamiltonian is
divergent.
\be \langle \psi_k | H_0 | \psi_k \rangle  \rightarrow  \infty \;.
\ee
It is a well known fact that point interactions on $\mathbb{R}^D$
can be considered as a self adjoint extension of the free
Hamiltonian \cite{Albeverio 2004, Jackiw}. We may heuristically
think of our problem as a kind of self adjoint extension since the
wave function $\psi_k(x)$ that we have found does not belong to
the domain of the free Hamiltonian so self adjoint extension of
the free Hamiltonian extends the domain of it such that the states
corresponding to the eigenfunctions $\psi_k(x)$ are included.

\section{Lower Bound of the Ground State Energy $ E_{gr}$}
\label{lowerbound gse delta}

Although we renormalize the model, we have not completely proven
that the energy is bounded from below. A well-known theorem in
matrix analysis, called Ger\v{s}gorin theorem \cite{horn} states
that all the eigenvalues $\omega$ of a matrix $\Phi \in M_N$ are
located in the union of $N$ discs
\be \label{gerstheorem} \bigcup_{i=1}^{N} \{|\omega-
\Phi_{ii}|\leq R'_{i}(\Phi)\} \equiv G(\Phi) \;, \ee
where $R'_{i}(\Phi)\equiv \sum_{i \neq j =1}^{N} |\Phi_{ij}| $ is
the deleted absolute row sums. Since the matrix $\Phi_{ij}(E)$ is
Hermitian due to the symmetry property of heat kernel, we have
$\omega \in \mathbb{R}$. Indeed, all eigenvalues $\omega$ are zero
in our problem. If there is a lower bound on energy, that is,
ground state energy, then we must expect that there would be no
solution at all beyond this lower bound, say $E_{*}=-\nu_{*}^{2}$.
Then, we want $\omega=0$ not to be an eigenvalue, thereby none of
the discs defined above should contain the zero eigenvalue when $E
\leq E_{*}$. This means that we should impose
\be \label{gersh} |- \Phi_{ii}(E)|= \Phi_{ii}(E)
> \sum_{i \neq j}^{N}|\Phi_{ij}(E)| \;, \ee
for all $i$, that is, \textit{the principal matrix must be
strictly diagonally dominant in order not to have a zero
eigenvalue.} However, before imposing this condition we can
simplify the problem. We note that
\beqs |\Phi_{ii}(E)| & \geq & |\Phi_{ii}(E)|^{\min} \cr \sum_{i
\neq j}^{N}|\Phi_{ij}(E)| & \leq & (N-1) |\Phi_{ij}(E)|^{\max} \;,
\eeqs
so the above condition (\ref{gersh}) is implied by the stronger
requirement
\be |\Phi_{ii}(E)|^{\min} >  (N-1) |\Phi_{ij}(E)|^{\max}\;.
\label{lowerbound condition delta} \ee
Once we obtain a solution to this inequality, it is satisfied for
all $E \leq E_{*}$ since the diagonal part of the principal matrix
(\ref{principal operator}) is a decreasing function of $E$ and the
off-diagonal part of it is an increasing function for given $a_i$,
$a_j$ and $N$. This means that there is no solution beyond this
critical value $E_{*}$. Hence, the ground state energy must be
larger than the critical value $E_{*}$:
\be E_{gr} \geq E_{*} \;. \ee
The basic idea of the proof was given for special manifolds
$\mathbb{S}^2$, $\mathbb{H}^2$ and $\mathbb{H}^3$ in our previous
work \cite{erman}. In the next subsections, we will find the lower
bound of the ground state energy for more general class of
manifolds.

\subsection{Compact Manifolds}
\label{The Proof for Compact Manifolds}

For compact manifolds, the upper bound for the off-diagonal
elements of heat kernel  and the lower bound for on-diagonal part
of the heat kernel has been given in (\ref{compactoffdiagupp}) and
(\ref{compactdiaglow}), respectively. Using these bound estimates
in (\ref{principal operator}), we find a lower bound for the
principal matrix
\be \Phi_{ii} (E) \geq
\begin{cases}
\begin{split}
{1 \over \left(4 \pi \hbar^2 /2m \right)} \ln\left( \nu^2/\mu^2
\right)
\end{split}
& \textrm{if $D = 2$} \\
\begin{split}
{2\sqrt{\pi} \over \hbar \left(4 \pi \hbar /2m \right)^{3/2}}
\left( \sqrt{{\nu^2 \over \hbar}} - \sqrt{{\mu^2 \over \hbar}}
\right)
\end{split}
& \textrm{if $D = 3$}\;,
\end{cases}
\label{lower bound diagonal phi compact} \ee
and an upper bound for it
\be |\Phi_{ij}(E)| \leq
\begin{cases}
\begin{split}
4A \Bigg[ \sqrt{{2 \over C_2}} { K_1\left( \sqrt{{2 \over C_2}}
{\nu \over \mu_d}\right) \over V(\mathcal{M})\nu \mu_d}  + {2
B(\varepsilon) K_0\left(\sqrt{{2 \over C_2}} {\nu \over
\mu_d}\right) \over \left(4\pi \hbar^2/2m \right)} \Bigg]
\end{split}
& \textrm{if $D = 2$} \\
\begin{split}
4A \Bigg[ \sqrt{{2 \over C_2}} { K_1\left( \sqrt{{2 \over C_2}}
{\nu \over \mu_d}\right) \over V(\mathcal{M})\nu \mu_d}  + {
\sqrt{2\pi C_2} B(\varepsilon) \mu_d \exp\left( - \sqrt{{2 \over
C_2}} {\nu \over \mu_d}\right) \over \hbar^{3/2} \left(4\pi
\hbar/2m \right)^{3/2}} \Bigg]
\end{split}
& \textrm{if $D = 3$}\;,
\end{cases}
\ee
where $i\neq j$ and we have used the monotonic behavior of the
functions in $\Phi_{ij}(E)$ so that we could maximize the
principal matrix in which we defined $d \equiv \min_{i\neq j}
d_{ij}$ and $\mu \equiv \max_{i}\mu_i$. We also introduced a
natural energy scale $\mu_d^2 \equiv {\hbar^2 \over 2m d^2}$ for
simplicity. In order to solve the inequality analytically, we must
estimate the bounds on the Bessel and logarithm functions. We can
estimate the lower bound of logarithmic function \cite{Abramowitz}
\be \label{logestimates}\ln u > \frac{u-1}{u} \hspace{1cm}
\mathrm{for}\; u > 0 \;,u \neq 1 \;. \ee
Let us first consider two dimensional case. As a result of bounds
(\ref{bessel0estiamte}), (\ref{bessel1estimates}) given for Bessel
functions and the one for logarithmic function given above, we
find
\be \Phi_{ii} (E)\geq {2 m \over \pi \hbar^2} \left(1-{1 \over
\nu/\mu}\right) \;,\ee
and
\begin{eqnarray} |\Phi_{ij} (E)| & \leq & 4A\exp
\left(- \sqrt{{1 \over 2 C_2}} {\nu \over \mu_d} \right)  \Bigg[
{1 \over V(\mathcal{M})} \Bigg( {1 \over \nu^2} + {1 \over \sqrt{2
C_2}\nu \mu_d} \Bigg) \cr &+& {2\sqrt{2C_2} B(\varepsilon) \over
\left(4\pi \hbar^2/2m \right) \nu/\mu_d}\Bigg] \;.
\end{eqnarray}
Since $\nu > \mu_d$, we may have
\beqs |\Phi_{ij} (E)| & \leq & 4A { \exp\left(- \sqrt{{1 \over 2
C_2}} {\nu \over \mu_d} \right) \over \nu/\mu_d} \cr & &
\hspace{2cm} \times \Bigg[ {1 \over V(\mathcal{M}) \mu_d^2} \Bigg(
1 + {1 \over \sqrt{2C_2}} \Bigg) + {2\sqrt{2C_2}B(\varepsilon)
\over \left(4\pi \hbar^2/2m \right)}\Bigg] \;. \eeqs
Therefore, there will be no solution to the eigenvalue equation
for values of the ground state energy below a critical value
$\nu>\nu_{*}$ if the following inequality is satisfied
\begin{eqnarray}  {2 m \over \pi
\hbar^2} \left(1-{1 \over \nu/\mu}\right) & > & 4A(N-1){
\exp\left(- \sqrt{{1 \over 2 C_2}} {\nu \over \mu_d} \right) \over
\nu/\mu_d} \Bigg[ {1 \over V(\mathcal{M})\mu_d^2} \Bigg( 1 + {1
\over \sqrt{2C_2}} \Bigg) \cr & + & {2\sqrt{2 C_2}B(\varepsilon)
\over \left(4\pi \hbar^2/2m \right)}\Bigg] \;.
\end{eqnarray}
Hence, we can solve this and get the lower bound for the ground
state energy
\beqs  E_{gr}  \geq - \nu_{*}^{2}= - \mu_{d}^{2} \Bigg[{\mu \over
\mu_d} + \sqrt{2C_2} \; W \bigg((N-1) A_1 \bigg) \Bigg]^2 \;,
\eeqs
where
\begin{eqnarray}
A_1 \equiv  {\exp\left(- \sqrt{{1 \over 2 C_2}} {\mu \over \mu_d}
\right) \over \sqrt{2C_2}} \Bigg[ {1 \over V(\mathcal{M})\mu_d^2}
\Bigg( 1 + {1 \over \sqrt{2C_2}} \Bigg)  +
{2\sqrt{2C_2}B(\varepsilon) \over \left(4\pi \hbar^2/2m
\right)}\Bigg] 4A \left(\pi \hbar^2/2m \right)\;,
\end{eqnarray}
and $W$ is called Lambert-W function, also called the Omega or the
ProductLog function \cite{Corless} and it is defined as the
inverse function of $x e^{x}$. In other words,
\be y=x e^{x} \Longleftrightarrow x = W(y) \;. \label{w function}
\ee
As for the three dimensional case, the off-diagonal part of the
principal matrix have the following upper bound
\begin{eqnarray}\hskip-0.5cm |\Phi_{ij} (E)| & \leq & 4A \exp\left(- \sqrt{{1 \over
2 C_2}} {\nu \over \mu_d} \right) \Bigg[ {1 \over
V(\mathcal{M})\mu_d^2} \Bigg( 1 + {1 \over \sqrt{2C_2}} \Bigg)\cr
&+&  {\sqrt{2\pi C_2} B(\varepsilon) \mu_d \over \hbar \left(4\pi
\hbar^2/2m \right)}\Bigg] \;,
\end{eqnarray}
where we use less sharp upper bound by using $\nu > \mu_d$ in
order to solve the inequality. Therefore, we conclude that there
exists a critical value of bound state energy $\nu>\nu_{*}$ for a
given $N$ and $d$ such that $\omega \neq 0$ so the ground state
energy cannot be less than $-\nu_{*}^{2}$
\begin{eqnarray}  E_{gr} & \geq & - \nu_{*}^{2}= - \mu_d^2 \Bigg[ {\mu \over \mu_d} + \sqrt{{C_2 \over 2}}\; W
\left((N-1) A_2 \right) \Bigg]^2 \;,
\end{eqnarray}
where
\begin{eqnarray}
A_2 & \equiv & {\exp\left(- \sqrt{{2 \over C_2}} {\mu \over \mu_d}
\right) \over \sqrt{2 \pi C_2}} \Bigg[ {1 \over
V(\mathcal{M})\mu_d^2} \Bigg( 1 + {1 \over \sqrt{2C_2}} \Bigg) +
{\sqrt{2 \pi C_2}B(\varepsilon) \mu_d \over \hbar^{3/2} \left(4\pi
\hbar/2m \right)^{3/2}}\Bigg] \cr & \times & {\hbar^{3/2} 4A
\left(4 \pi \hbar/2m \right)^{3/2} \over \mu_d} \;.
\end{eqnarray}

\subsection{Cartan-Hadamard Manifolds}
\label{The Proof for Cartan-Hadamard Manifolds}

Similarly, using the upper and lower bounds of the heat kernel for
Cartan-Hadamard manifolds, we have obtained the upper and lower
bound on the off and on diagonal part of the principal matrix,
respectively:
\be \Phi_{ii} (E) \geq
\begin{cases}
\begin{split}
{c \over \left(4 \pi \hbar^2 /2m \right)} \ln\left( { {\nu^2 \over
\hbar} + \xi \over {\mu^2 \over \hbar} + \xi } \right)
\end{split}
& \textrm{if $D = 2$} \\
\begin{split}
{2\sqrt{\pi} c \over \hbar \left(4 \pi \hbar /2m \right)^{3/2}}
\left( \sqrt{{\nu^2 \over \hbar} + \xi} - \sqrt{{\mu^2 \over
\hbar} +\xi} \right)
\end{split}
& \textrm{if $D = 3$} \;,
\end{cases}
\label{lower bound diagonal phi CH} \ee
where $\xi\equiv {\hbar (\sigma_1(\mathcal{M})+\delta) \over 2m}
\geq 0$ and
\be \Phi_{ij} (E) \leq
\begin{cases}
\begin{split}
{2 C(\varepsilon,\kappa) \over \left(4 \pi \hbar^2 /2m \right)}
K_0\left(\sqrt{{2 \over C_2}} {\nu \over \mu_d} \right)
\end{split}
& \textrm{if $D = 2$} \\
\begin{split}
{\sqrt{2 \pi C_2} C(\varepsilon,\kappa) \mu_d \over \hbar^{3/2}
\left(4 \pi \hbar /2m \right)^{3/2}} \exp\left(- \sqrt{{2 \over
C_2}} {\nu \over \mu_d} \right)
\end{split}
& \textrm{if $D = 3$} \;,
\end{cases}
\ee
for $i \neq j$. For $D=2$, we have
\be \ln\left( { {\nu^2 \over \hbar} + \xi \over {\mu^2 \over
\hbar} + \xi } \right) \geq \ln \left( { \xi \over {\mu^2 \over
\hbar} + \xi } \right)\ee
and if we use the same bounds for the Bessel function and the fact
that $\nu>\mu_d$, we obtain the lower bound for the ground state
energy
\be E_{gr} \geq - 2 C_2 \mu_d^{2}\; W^2 \left( {2 (N-1)
C(\varepsilon,\kappa) \over \ln\left({\xi \over {\mu^2 \over
\hbar}+\xi}\right)} \right) \;. \ee
For three dimensional case, we must do some additional assumption
in order to get an analytic solution. We will assume that $\nu
\geq \mu^2 + \hbar\xi$ and this assumption should be checked
whether it is consistent or not after we have found the solution.
Hence,
\be \left( \sqrt{{\nu^2 \over \hbar} + \xi} - \sqrt{{\mu^2 \over
\hbar} +\xi} \right)
 \geq  \left(  {\nu \over \hbar^{1/2}} - \sqrt{{\mu^2 \over \hbar} +\xi}
\right) \;, \ee
and finally
\be E_{gr}\geq - \mu_d^2 \left[{1 \over \mu_d} \sqrt{\mu^2 + \hbar
\xi}
 + \sqrt{{C_2 \over 2}} W \left(A_3
(N-1)\right)\right]^2 \;, \ee
where
\be A_3 \equiv {C(\varepsilon,\kappa) \over c} \exp\left(- {1
\over \mu_d} \sqrt{{2 \over C_2}\left(\mu^2 + \hbar \xi \right)}
\right) \;.\ee

\section{Non-degeneracy and Positivity of the Ground State}
\label{Non-degeneracy and Positivity of the Ground State}

The rigorous proof of non-degeneracy and positivity of the ground
state in standard quantum mechanics is given in
\cite{simon,Berezin Shubin}, which does not include the singular
potentials. Therefore, it is necessary to check whether this is
still valid for our problem. The proof in our case is based on
Perron - Frobenius theorem \cite{horn}: It states that If $A\in
M_N$ and if we suppose that $A>0$ (that is, all $A_{ij} >0$), then

(a) $\rho(A)>0$;

(b) $\rho(A)$ is an eigenvalue of $A$;

(c) There is an $x\in \mathbb{C}^{N}$ with $x>0$ and $A x =
\rho(A)x$;

(d) $\rho(A)$ is an algebraically (and hence geometrically) simple
eigenvalue of $A$;

(e) $|\omega|< \rho(A)$ for every eigenvalue $\omega \neq
\rho(A)$, that is, $\rho(A)$ is the unique eigenvalue of maximum
modulus. Here $\rho(A)=\max\{|\omega|: \omega \; \; \mathrm{is}
\;\; \mathrm{an} \;\; \mathrm{eigenvalue} \;\; \mathrm{of} \;\;
A\}$ and called spectral radius. A simple proof of this theorem
for the positive symmetric matrices is given in \cite{Ninio}.

Since the principal matrix (\ref{principal operator}) is not a
positive matrix, we cannot directly apply Perron-Frobenius
theorem. Nevertheless, we can make the principal matrix strictly
positive by subtracting the maximum of the diagonal part of it
corresponding to the lower bound of energy $E=E_{*}$, which is
found in section \ref{lowerbound gse delta}, and reversing the
overall sign:
\be \Phi'(E)= -\left[\Phi(E)- (1+\varepsilon) \, \mathbf{1} \,
\Phi^{\max}_{ii}(E_{*})\right] > 0 \;, \hspace{1cm} \varepsilon>0
\;, \ee
where $\mathbf{1}$ is a $N\times N$ identity matrix. Hence,
considering the transformed principal matrix $\Phi'$ in the light
of this theorem, we conclude that there exist a strictly positive
eigenvector which corresponds to the unique eigenvalue of maximum
modulus.
\be \sum_{j=1}^{N} \Phi'_{ij}(E) A_{j}(E)= \rho(\Phi') A_{i}(E)
\;. \ee
Here it must be noticed that $\Phi'$ has the same eigenvector with
$\Phi$ so it guarantees that there exist a strictly positive
eigenvector for the principal matrix $\Phi$. Using the eigenvalue
problem, we find
\be \sum_{j=1}^{N} \Phi_{ij}(E) A_j(E) = \omega(E) A_i(E) \;, \ee
where $\rho(\Phi')= -\omega^{\min}(E)+(1+\varepsilon)
\Phi_{ii}(E_{*})$. For a given $E=E_k$ or $\nu=\nu_k$, there is a
unique corresponding $\omega^{\min}(E)$ and since we are looking
for the zeros of the eigenvalues $\omega(E)=0$, this minimum flows
to zero at $\nu=\nu^{\max}=\nu_{*}$. This means that the positive
eigenvector $A_i$ corresponds to the ground state energy so we
prove that the ground state energy is unique and the associated
eigenvector $A_i$ is strictly positive. Due to the positivity
property of heat kernel, it is easy to see that the ground state
wave function is strictly positive from the equation
(\ref{wavefunction heat delta}).
\beqs \psi_k(x) & = & \left[ \sum_{i,j=1}^N
\underbrace{A_i(-\nu_k^2)}_{> 0}
 \int_{0}^{\infty} {\mathrm{d} t \over \hbar} \;
\underbrace{\left({t \over \hbar} \right) K_t(a_i,a_j;g)
e^{-\frac{t \nu_{k}^2}{\hbar}}}_{\geq 0} \,
\underbrace{A_j(-\nu_k^2)}_{>0} \right]^{-\frac{1}{2}} \; \cr & &
\hspace{2cm}\times \int_{0}^{\infty} {\mathrm{d} t  \over \hbar}\;
\underbrace{e^{-\frac{t \nu_{k}^2}{\hbar}}}_{> 0} \sum_{i=1}^N
\underbrace{A_i(-\nu_k^2)}_{>0} \underbrace{K_{t}(a_i, x;g)}_{> 0}
> 0 \;, \eeqs
Hence, we prove that despite the singular character of the
interaction, the ground state is still non-degenerate and unique.

\section{Renormalization Group Equations}
\label{Renormalization Group}

Renormalization group equations of our problem (for $N=1$ case) in
flat spaces has been already given in the literature
\cite{Tarrach2,Adhikari,Camblong2}. Here, we will show that this
can be also derived explicitly for our problem.

\subsection{Two Dimensional Case}
\label{rg 2d}

\textit{In this section, we shall choose the natural units
$\hbar=2m=1$ for simplicity}. It is useful to work with the
dimensionless coupling constant. One possible way for the
renormalization scheme in order to determine how the coupling
constant changes with the energy scale is to define the following
renormalized coupling constant $\lambda^{R}_{i}(M_i)$ in terms of
the bare coupling constant $\lambda_i(\epsilon)$
\be {1 \over \lambda^{R}_{i}(M_i)} = {1 \over
\lambda_{i}(\epsilon)} - \int_{\epsilon}^{\infty} \mathrm{d} t
{e^{- M_{i}^{2} t} \over 4 \pi t} \;,  \ee
where $M_i$ is the renormalization scale (it is of dimension
$[E]^{1/2}$). Then, the renormalized principal matrix in terms of
renormalized coupling constant in natural units is given
\be \Phi_{ij}^{R} (E) =
\begin{cases}
\begin{split} {1 \over \lambda_{i}^{R}(M_i)}-
\int_0 ^\infty \mathrm{d} t  \left( K_{t}(a_i,a_i;g) e^{ t E} -
{e^{- M_{i}^{2} t} \over 4 \pi t}\right)
\end{split}
& \textrm{if $i = j$} \\
\begin{split}
- \; \int_0^\infty \mathrm{d} t \;  K_{t}(a_i,a_j;g)e^{t E}
\end{split}
& \textrm{if $i \neq j$} \;,
\end{cases}
\ee
and the bound state energy is determined from the condition $\det
\Phi_{ij}^{R} (E) =0$ and it determines the relation between
$\lambda_{i}^{R}(M_i)$ and $M_i$. Explicit dependence on $M_i$
cancels the implicit dependence on $M_i$ through
$\lambda_{i}^{R}(M_i)$. Physics is determined by the value of
$\lambda_{i}^{R}(M_i)$ at an arbitrary value of the
renormalization point $M_i$. However, this is not a proper way to
look at our problem since we have to deal with several
renormalized coupling constants with the same kind of interaction,
which essentially differs from one another with arbitrary
constants. These constants can be determined by deciding the
excited energy levels. We shall instead prefer one renormalized
coupling constant by redefining the meaning of the renormalized
coupling constant without altering physics. This could be done in
the following way: As an external input, we decide about the
relative strengths of individual delta interactions and do not use
the ground state energy to fix the flow. We know that
$-\mu_{i}^{2}$ is the bound state energy of the individual $i$ th
Dirac delta center so it corresponds to the solution
$\Phi_{ii}^{R}(-\mu_{i}^{2})=0$. Without loss of generality, let
us assume that $\Phi_{11}^{R}(-\mu_{1}^{2})=0$ and this allows us
to choose the renormalized coupling constant
\be {1 \over \lambda^{R}(M)} = {1 \over \lambda_{1}(\epsilon)} -
\int_{\epsilon}^{\infty} \mathrm{d} t  {e^{- M^{2} t} \over 4 \pi
 t } \label{renormcoupling 1} \;, \ee
at some scale $M$. Once the renormalized coupling constant is
fixed under this condition, we must also satisfy
$\Phi_{ii}^{R}(-\mu_{i}^{2})=0$ for $i\neq 1$ with this choice at
the same scale $M$. This is always possible if we add a constant
term to the definition of renormalized coupling constant. Let us
consider $i=2$ case
\beqs \Phi_{22}^{R}(-\mu_{2}^{2}) &=& {1 \over \lambda^{R}(M)} +
\lim_{\epsilon \rightarrow 0} \Bigg[ \int_{\epsilon}^{\infty}
\mathrm{d} t \;  {e^{- M^{2} t} \over 4 \pi
 t} -  \int_\epsilon^\infty \mathrm{d} t \;
K_{t}(a_2,a_2;g) e^{- t \mu_{2}^{2}} \Bigg]- \Sigma_2 \cr &=&
\int_0 ^\infty \mathrm{d} t \;  \Bigg[ K_{t}(a_1,a_1;g) e^{- t
\mu_{1}^{2}}-  K_{t}(a_2,a_2;g) e^{- t \mu_{2}^{2}} \Bigg] -
\Sigma_2 =0 \;, \eeqs
where we have used (\ref{renormcoupling 1}) and
$\Phi_{11}^{R}(-\mu_{1}^{2})=0$. This means that there always
exist a constant $\Sigma_i$ depending only on $\mu_i$ with
$\Sigma_1=0$ and $\Sigma_i \neq 0$ for $i\neq 1$ such that the
condition $\Phi_{ii}^{R}(-\mu_{i}^{2})=0$ can be satisfied. Hence,
the renormalized coupling constant becomes
\be {1 \over \lambda^{R}(M)} = {1 \over \lambda_{i}(\epsilon)} -
\int_{\epsilon}^{\infty} \mathrm{d} t \; {e^{- M^{2} t} \over 4
\pi
 t} + \Sigma_i \label{renormcoupling 2d} \;, \ee
and the choice of $\Sigma_i$ s refer to the relative strengths of
delta interactions in this renormalization scheme. As a result,
the renormalized principal matrix is
\be \Phi_{ij}^{R} (E) =
\begin{cases}
\begin{split} {1 \over \lambda_{R}(M)}-
\int_0 ^\infty \mathrm{d} t \; \left( K_{t}(a_i,a_i;g) e^{ t E} -
{e^{- M^2 t} \over 4 \pi t}\right)-\Sigma_i
\end{split}
& \textrm{if $i = j$} \\
\begin{split}
- \; \int_0^\infty \mathrm{d} t \; K_{t}(a_i,a_j;g)e^{t E}
\end{split}
& \textrm{if $i \neq j$}.
\end{cases} \label{renormalized principal 2d}
\ee
The renormalization condition is given by
\be M {\mathrm{d} \Phi_{ij}^{R}(M,\lambda_R(M),E;g) \over
\mathrm{d} M} =0 \;,\label{renorm cond} \ee
or
\be \left( M {\partial \over \partial M} + \beta(\lambda_R)
{\partial \over \partial \lambda_R}
\right)\Phi_{ij}^{R}(M,\lambda_R(M),E;g)=0 \;,\label{renorm
condition 2d} \ee
where
\be \label{beta function def} \beta(\lambda_R)= M {\partial
\lambda_R \over
\partial M} \ee
is called $\beta$ function and the equation (\ref{renorm condition
2d}) is the renormalization group (RG) equation. In
\cite{Adhikari}, the renormalization condition (\ref{renorm cond})
corresponding to the problem in flat space has been written in
terms of the $T$-matrix. Using (\ref{renormalized principal 2d})
in (\ref{renorm condition 2d}), we can find $\beta$ function
exactly
\be \beta(\lambda_R) = -{\lambda_{R}^{2} \over 2 \pi} < 0 \;. \ee
This result is the same as the one in flat spaces given in the
literature \cite{Tarrach2,Nyeo,Adhikari,Camblong2} so our problem
is asymptotically free. From the explicit expression of the
renormalized principal matrix, one can easily see the scaling
property of it under a change of energy and metric scale $\gamma$
using the scaling property of the heat kernel
(\ref{heatkernelscaling2})
\be \Phi_{ij}^{R}(M, \lambda_R(M), \gamma^2 E;\gamma^{-2}g) =
\Phi_{ij}^{R}(\gamma^{-1} M, \lambda_R(M), E;g) \;. \ee
It is important to notice that we need to scale the metric as well
and the idea of the metric scaling in deriving the renormalization
group equation was motivated by \cite{odintsov} in the context of
renormalization group in quantum field theory on curved spaces.
Hence we have
\be \gamma {\mathrm{d} \over \mathrm{d} \gamma}
\left[\Phi_{ij}^{R}(M, \lambda_R(M), \gamma^2 E;\gamma^{-2}g) =
\Phi_{ij}^{R}(\gamma^{-1} M, \lambda_R(M), E; g) \right] \;. \ee
This leads to the renormalization group equation for
$\Phi_{ij}^{R}(M, \lambda_R(M), \gamma^2 E;\gamma^{-2}g)$
\be \gamma {\mathrm{d} \over \mathrm{d} \gamma} \Phi_{ij}^{R}(M,
\lambda_R(M), \gamma^2 E;\gamma^{-2}g) + M {\partial \over
\partial M} \Phi_{ij}^{R}(M, \lambda_R(M), \gamma^2
E;\gamma^{-2}g) =0 \;, \ee
or
\be \left[\gamma {\mathrm{d} \over \mathrm{d} \gamma} -
\beta(\lambda_R) {\partial \over
\partial \lambda_R}\right] \Phi_{ij}^{R}(M, \lambda_R(M), \gamma^2
E;\gamma^{-2}g) =0 \;. \label{rg equation 2d} \ee
If we postulate the following functional form for the principal
matrix
\be \Phi_{ij}^{R}(M, \lambda_R(M), \gamma^2 E;\gamma^{-2}g) =
f(\gamma) \Phi_{ij}^{R}(M, \lambda_R(\gamma M), E;g)
\label{funtional ansatz} \;,\ee
and substitute into (\ref{rg equation 2d}) we obtain an ordinary
differential equation for the function $f$
\be  \gamma {\mathrm{d} f(\gamma) \over \mathrm{d} \gamma} =0 \;.
\ee
This gives the solution $f(\gamma)=1$ using the initial condition
at $\gamma=1$. Therefore, we get
\be \Phi_{ij}^{R}(M, \lambda_R(M), \gamma^2 E;\gamma^{-2}g) =
\Phi_{ij}^{R}(M, \lambda_R(\gamma M), E;g) \;, \label{grm scaling
2d} \ee
which means that there is no anamolous scaling. After integrating
\be \beta(\lambda_R) = \bar{M} {\partial \lambda_R(\bar{M})\over
\partial \bar{M}} = -{\lambda_{R}^{2}(\bar{M}) \over 2 \pi} \ee
between  $\bar{M}= M$ to $\bar{M}=\gamma M$ we can find the flow
equation for the coupling constant
\be \lambda_R(\gamma M) = {\lambda_R(M) \over 1+ {1 \over 2 \pi}
\lambda_R(M)
 \ln \gamma} \;. \label{coupling const evolve 2d} \ee
One can explicitly check the relation (\ref{grm scaling 2d}) if
the coupling constant evolves according to (\ref{coupling const
evolve 2d}). First, we add and subtract a term in the time
integral:
\beqs & & \hskip-1cm \Phi_{ii}^{R}(M, \lambda_R(\gamma M), E;g) =
{1 \over \lambda_{R}(M)} + {1 \over 2 \pi} \ln \gamma - \int_0
^\infty \mathrm{d} t \;  \Bigg( K_{t}(a_i,a_i;g) e^{ t E} - {e^{-
M^2 t } \over 4 \pi t}\Bigg) - \Sigma_i  \cr & & \hskip-1cm = {1
\over \lambda_{R}(M)} + {1 \over 2 \pi } \ln \gamma - \int_0
^\infty \mathrm{d} t \; \Bigg( K_{t}(a_i,a_i;g) e^{ t E}  - {e^{-
M^2 t} \over 4 \pi t} + {e^{- M^2 \gamma^{-2} t} \over 4 \pi t} -
{e^{- M^2 \gamma^{-2} t} \over 4 \pi t}\Bigg)-\Sigma_i \cr &=& {1
\over \lambda_{R}(M)}  - \int_0 ^\infty \mathrm{d} t \; \Bigg(
K_{t}(a_i,a_i;g) e^{ t E}  - {e^{- M^2 \gamma^{-2} t} \over 4 \pi
t}\Bigg)-\Sigma_i\eeqs
and then using the scaling property of heat kernel
(\ref{heatkernelscaling2}), we get
\beqs & &  {1 \over \lambda_{R}(M)}  - \int_0 ^\infty \mathrm{d} t
\; \Bigg( \gamma^{-2} K_{\gamma^{-2} t}(a_i,a_i;\gamma^{-2} g) e^{
t E} - {e^{- M^2 \gamma^{-2} t} \over 4 \pi t}\Bigg)-\Sigma_i \cr
&=&  {1 \over \lambda_{R}(M)}  - \int_0 ^\infty \mathrm{d} s \;
\Bigg( K_{s}(a_i,a_i;\gamma^{-2} g) e^{ s \gamma^2 E} - {e^{- M^2
s} \over 4 \pi s}\Bigg)-\Sigma_i \cr & = & \Phi_{ii}^{R}(M,
\lambda_R(M), \gamma^2 E;\gamma^{-2}g) \;. \eeqs
Off diagonal term can be directly checked using just the scaling
property of heat kernel (\ref{heatkernelscaling2}). In another way
of thinking, one can find how the coupling constant evolves, given
(\ref{coupling const evolve 2d}) from the scaling relation
(\ref{grm scaling 2d}).

\subsection{Three Dimensional Case}
\label{rg 3d}

Since it is convenient to work with the dimensionless coupling
constant, we define a dimensionless coupling constant in three
dimensions
\be \hat{\lambda}_R(M) = M \lambda_R(M) \;. \ee
Then, by similar arguments developed for two dimensions, the
renormalized principal matrix in the natural units is
\be \Phi_{ij}^{R} (E) =
\begin{cases}
\begin{split} {M \over \hat{\lambda}_{R}(M)}-
\int_0 ^\infty \mathrm{d} t \; \left( K_{t}(a_i,a_i;g) e^{ t E} -
{e^{- M^2 t } \over (4 \pi t)^{3/2}}\right)-\Sigma_i
\end{split}
& \textrm{if $i = j$} \\
\begin{split}
- \; \int_0^\infty \mathrm{d} t \; K_{t}(a_i,a_j;g)e^{t E}
\end{split}
& \textrm{if $i \neq j$}.
\end{cases} \label{renormalized principal 3d}
\ee
Renormalization condition (\ref{renorm condition 2d}) in this case
leads to the following $\beta$ function
\be \beta(\hat{\lambda}_R) = M {\partial \hat{\lambda}_R(M) \over
\partial M} = \hat{\lambda}_R(M) -{1 \over 4\pi}
\hat{\lambda}_{R}^{2}(M) \;, \ee
which is in agreement with the result for flat space
\cite{Adhikari}. From the explicit expression of the renormalized
principal matrix for three dimensions, one can easily see the
scaling property of it under a change of scale $\gamma$ using the
scaling property of the heat kernel (\ref{heatkernelscaling2})
\be \Phi_{ij}^{R}(M, \hat{\lambda}_R(M), \gamma^2 E;\gamma^{-2} g)
= \gamma \Phi_{ij}^{R}(\gamma^{-1} M, \hat{\lambda}_R(M), E; g)
\;, \ee
so we have
\be \gamma {\mathrm{d} \over \mathrm{d} \gamma}
\left[\Phi_{ij}^{R}(M, \hat{\lambda}_R(M), \gamma^2 E;\gamma^{-2}
g) = \gamma \Phi_{ij}^{R}(\gamma^{-1} M, \hat{\lambda}_R(M), E; g)
\right] \;. \ee
This leads to the renormalization group equation for
$\Phi_{ij}^{R}(M, \hat{\lambda}_R(M), \gamma^2 E;\gamma^{-2} g)$
\be \left[\gamma {\mathrm{d} \over \mathrm{d} \gamma}-1 + M
{\partial \over
\partial M} \right] \Phi_{ij}^{R}(M, \hat{\lambda}_R(M), \gamma^2
E;\gamma^{-2} g) =0 \;, \ee
or
\be \left[\gamma {\mathrm{d} \over \mathrm{d} \gamma} -1 -
\beta(\hat{\lambda}_R) {\partial \over
\partial \hat{\lambda}_R}\right] \Phi_{ij}^{R}(M, \hat{\lambda}_R(M), \gamma^2
E;\gamma^{-2} g) =0 \label{rg equation 3d} \;. \ee
If we again postulate (\ref{funtional ansatz}) and substitute it
into (\ref{rg equation 3d}), we obtain an ordinary differential
equation for the function $f$
\be  \gamma {\mathrm{d} f(\gamma) \over \mathrm{d} \gamma} =f \;.
\ee
The solution is $f(\gamma)=\gamma$ by using the initial condition
at $\gamma=1$. Therefore, we have
\be \Phi_{ij}^{R}(M, \hat{\lambda}_R(M), \gamma^2 E;\gamma^{-2}g)
= \gamma \Phi_{ij}^{R}(M, \hat{\lambda}_R(\gamma M), E; g)
\label{grm scaling 3d}\;. \ee
This means that there is also no anamolous scaling in three
dimensions. After integrating
\be \beta(\hat{\lambda}_R) = \bar{M} {\partial
\hat{\lambda}_R(\bar{M})\over
\partial \bar{M}} = \hat{\lambda}_R(\bar{M}) -{1 \over 4\pi}
\hat{\lambda}_{R}^{2}(\bar{M}) \;,\ee
between  $\bar{M}= M$ to $\bar{M}=\gamma M$ we can find similarly
the flow equation for the coupling constant
\be \hat{\lambda}_R(\gamma M) = {\gamma \hat{\lambda}_R(M) \over 1
- {1 \over 4\pi} \hat{\lambda}_R(M) (1-\gamma)} \;.
\label{coupling const evolve 3d} \ee
One can similarly check the relation (\ref{grm scaling 3d}) if the
coupling constant evolves according to (\ref{coupling const evolve
3d}). In this case, we have
\beqs & & \hskip-1cm \gamma \Phi_{ii}^{R}(M,
\hat{\lambda}_R(\gamma M), E; g)\cr & & \hskip-1cm  = {M \over
\hat{\lambda}_{R}(M)} + {1 \over 4\pi} M (1-\gamma)  -   \gamma
\int_0 ^\infty \mathrm{d} t \; \Bigg( K_{t}(a_i,a_i; g) e^{ t E} -
{e^{- M^2 t} \over (4 \pi t)^{3/2}}\Bigg) - \Sigma_i \cr & &
\hskip-1cm = {M \over \hat{\lambda}_{R}(M)} + {1 \over 4\pi} M
(1-\gamma) - \gamma \int_0 ^\infty \mathrm{d} t \; \Bigg(
K_{t}(a_i,a_i; g) e^{ t E} - {e^{- M^2 t} \over (4 \pi t)^{3/2}} +
{e^{- M^2 \gamma^{-2} t} \over (4 \pi t)^{3/2}} \cr & & - {e^{-
M^2 \gamma^{-2} t } \over (4 \pi t)^{3/2}}\Bigg)-\Sigma_i \cr & &
\hskip-1cm ={M \over \hat{\lambda}_{R}(M)} - \gamma \int_0 ^\infty
\mathrm{d} t \; \Bigg( K_{t}(a_i,a_i;g) e^{ t E}  - {e^{- M^2
\gamma^{-2} t} \over (4 \pi t)^{3/2}}\Bigg)-\Sigma_i \;, \eeqs
and then using the scaling property of heat kernel
(\ref{heatkernelscaling2}), we get
\beqs & &  {M \over \hat{\lambda}_{R}(M)}  - \gamma \int_0 ^\infty
\mathrm{d} t  \; \Bigg( \gamma^{-3} K_{\gamma^{-2}
t}(a_i,a_i;\gamma^{-2} g) e^{ t E} - {e^{- M^2 \gamma^{-2} t }
\over (4 \pi  t)^{3/2}}\Bigg)-\Sigma_i \cr &=& {M \over
\hat{\lambda}_{R}(M)}  - \int_0 ^\infty \mathrm{d} s \; \Bigg(
K_{s}(a_i,a_i;\gamma^{-2} g) e^{ s \gamma^2 E} - {e^{- M^2 s }
\over (4 \pi  s)^{3/2}}\Bigg)-\Sigma_i \cr & = & \Phi_{ii}^{R}(M,
\hat{\lambda}_R(M), \gamma^2 E;\gamma^{-2}g) \;. \eeqs
One can similarly find how the coupling constant evolves, given in
(\ref{coupling const evolve 3d}) from the scaling relation
(\ref{grm scaling 3d}).

\section{Conclusion}
\label{conclusion}

In this paper, we studied the bound state problem for several
Dirac delta interactions in various two and three dimensional
Riemannian manifolds. Our renormalization method is basically
inspired from \cite{rajeevbound} developed for many body version
of this problem on flat spaces. This method allows us to
renormalize the model non-perturbatively. It has been also shown
that the heat kernel plays a key role in the renormalization
procedure and help us to find lower bounds on the ground state
energy due to the sharp upper bound estimates on it for several
classes of manifolds. We proved that many well known theorems
given in standard quantum mechanics are still valid, such as
pointwise bounds on the wave functions, non-degeneracy and
uniqueness of the ground state although we have singular
interactions. Renormalization procedure does not change these
well-known results in standard quantum mechanics. Finally, we
studied the renromalization group equations and the $\beta$
function is exactly calculated.

\section{Acknowledgments}

\hspace{0.5cm} The authors gratefully acknowledge the many helpful
discussions with B. Altunkaynak, \c{C}. Dogan, O. A. Eden, K. S.
Gupta, B. Kaynak. O. T. T. 's research is partially supported by
the Turkish Academy of Sciences, in the framework of the Young
Scientist Award Program (OTT-TUBA-GEBIP/2002-1-18).

\section{Appendix A: Heat Kernel on Riemannian Manifolds}
\label{heat kernel}

Let $(\mathcal{M}, g)$ be a compact connected Riemannian manifold,
then there exists a complete orthonormal system of $C^\infty$
eigenfunctions $\{f_l \}_{l=0}^{\infty}$ in $L^2(\mathcal{M},
\mathrm{d}^{D}_{g}x)$ and the spectrum $\sigma(\mathcal{M},
g)=\{\sigma_l\} = \{0 = \sigma_0 \leq \sigma_1 \leq \sigma_2 \leq
\dots\}$, with $\sigma_l$ tending to infinity as $l \rightarrow
\infty$ and each eigenvalue has finite multiplicity: $-\nabla_g^2
f_l(x;g)=\sigma_l f_l(x;g)$ . The multiplicity of the first
eigenvalue $\sigma_0=0$ is one and corresponding eigenfunction is
constant and given by $f_0(x;g)=1/\sqrt{V(\mathcal{M})}$, where
$V(\mathcal{M})$ is the manifold $(\mathcal{M}, g)$. This theorem
is also called Hodge theorem for functions or spectral theorem
\cite{chavel2,Rosenberg}. This theorem is valid for
 Neumann, Dirichlet and mixed eigenvalue problems except for
$\sigma_0 > 0$, provided that the appropriate boundary condition
is imposed.

The operator $-\nabla_g^2$ is formally self-adjoint with respect
to the $L^2(\mathcal{M}, \mathrm{d}^{D}_{g}x)$, that is, the inner
product is defined as
\be (\psi_1,\psi_2)=\int_{\mathcal{M}} \mathrm{d}^{D}_{g}x \;
\psi_{1}^{*}(x) \psi_2(x) \;. \ee
Since it will be necessary for our purposes to know how the
eigenfunctions change under the scaling transformations in the
metric, we will use a notation which specifies the metric
structure of the functions, such as $f_l(x;g)$. The spectral
theorem provides us with all the tools of Fourier analysis, so
that we can expand any function $\psi(x)\in L^2(\mathcal{M},
\mathrm{d}^{D}_{g}x)$ in terms of the complete orthonormal
eigenfunctions $f_l(x;g)$
\be \label{expansionpsi} \psi(x) = \sum_{l \geq 0} \left(\psi(x),
f_l(x;g)\right) \,f_l(x;g) =\sum_{l \geq 0} C_l \,f_l(x;g)\;, \ee
Orthogonality and completeness of the eigenfunctions on compact
manifolds are
\beqs \delta_{k l}& = & \int_{\mathcal{M}} \mathrm{d}_{g}^{D} x \;
 f_{k}^{*}(x;g) f_l(x;g) \;,
 \cr  \delta^{D}_{g}(x,a_i) & = & \sum_{l \geq 0} f_l(x;g)
f_{l}^*(a_i;g), \label{expansion for compact} \eeqs
where $C_l$'s are expansion coefficients. $\delta^{D}_{g}(x,a_i)$
is the $D$ - dimensional normalized delta function at point $a_i
\in \mathcal{M}$,
\be \int_{\mathcal{M}} \mathrm{d}_{g}^{D} x \;
\delta^{D}_{g}(x,a_i) =1 \;. \label{delta normalization} \ee
Note that extra labels in the eigenfunction expansion must be
taken into account if the problem admits degeneracy.

Moreover, one may extend heuristically the problem on to some
noncompact manifolds in such a way that spectral theorem is still
applied. The relations such as completeness and orthogonality
relations are defined by an appropriate generalization of the
measures to the continuous distributions in the sense of
\cite{berezansky}. Since the spectrum does not have to be discrete
for noncompact manifolds, we may have in general
\beqs \psi(x) & =& \int \mathrm{d} \mu(l) \; \psi(l) \,f_l(x;g)\;,
\cr \delta_{k l}& = & \int_{\mathcal{M}} \mathrm{d}_{g}^{D} x \;
f_{k}^{*}(x;g) f_l(x;g) \;,
 \cr  \delta^{D}_{g}(x,a_i) & = & \int \mathrm{d} \mu(l)\; f_l(x;g)
f_{l}^*(a_i;g), \label{expansion for noncompact} \eeqs
where $d\mu(l)$ is the spectral measure and it includes continuous
spectrum as well as point spectrum. This should be taken with a
grain of salt and one must not forget that we may not have a
rigorous proof of the spectral theorem for some special noncompact
manifolds. From now on,  we assume that we are dealing with
manifolds which do not have such pathologies, that is, the
spectral theorem is applicable.

Although the notion of heat kernel can be defined on any
Riemannian manifold, the explicit formulas only exist for some
special class of manifolds, for example, Euclidean spaces
$\mathbb{R}^{D}$ \cite{grigoryan} and hyperbolic spaces
$\mathbb{H}^D$ \cite{heatkernelh3}. We will list some of the well
known properties of the heat kernel on any $(\mathcal{M}, g)$, and
give the short time asymptotic expansion of the heat kernel. Also
we derive the scaling property of the heat kernel. They are all
needed in our calculations.

1) Heat Equation: It satisfies heat equation since it is a
fundamental solution to it by definition.
\be \label{heatkernelequation} \hbar {\partial K_{t}(x,y;g) \over
\partial t}-{\hbar^2 \over 2m }\nabla^2_g K_{t}(x,y;g) = 0 \;, \ee
where $\nabla^2_g$ acts on the $x$ variable.

2) Initial condition: It solves the Cauchy problem.
\be \label{initialcon} \lim_{t\to 0^+} K_{t}(.,y;g)= \delta_g(.,y)
\;. \ee

3) Semi-group Property:
\be \label{semigroupprop} \int_{\mathcal{M}} \mathrm{d}^{D}_g z \;
K_{ t_1}(x,z;g)K_{ t_2}(z,y;g) = K_{t_1 + t_2}(x,y;g) \;. \ee

4) Symmetry Property:
\be \label{symmetryprop} K_{t} (x,y;g) = K_{t} (y,x;g) \;. \ee

5) Positivity Property:
\be \label{positivityprop} K_{t}(x,y;g) >  0 \;\; \; \mathrm{for}
\; \mathrm{all} \; t
>0 \;. \ee

If $\mathcal{M}$ is compact, then we have
\be K_{t}(x,y;g)= \sum_{l\geq 0} e^{-{\hbar t \over 2m} \sigma_l}
f_{l}^{*}(x;g) f_l(y;g)\; , \label{expheat} \ee
which converges uniformly on $\mathcal{M} \times \mathcal{M}$. The
analog of this sum on noncompact manifolds can be given as
\be K_{t}(x,y;g)= \int \mathrm{d} \mu(l) e^{-{\hbar t \over 2m}
\sigma(l)} f_{l}^{*}(x;g) f_l(y;g)\; . \label{expheatnoncompact}
\ee

We have also eigenfunction expansion of the creation and
annihilation operators
\be \phi_{g}^{\dagger}(x)= \sum_{l\geq 0} \phi_{g}^{\dagger}(l)
f_{l}^{*}(x;g)\; \label{eigenfuncexp} \ee
or in noncompact manifolds
\be \phi_{g}^{\dagger}(x)=  \int \mathrm{d} \mu(l) \;
\phi_{g}^{\dagger}(l) f_{l}^{*}(x;g)\;
\label{eigenfuncexpnoncompact} \ee
When the manifold $\mathcal{M}$ is a complete Riemannian manifold
with Ricci curvature bounded from below then the heat kernel
satisfies the stochastic completeness property \cite{grigoryan,
yau}:
\be \nonumber \int_{\mathcal{M}} \mathrm{d}_{g}^{D} x \;
K_{t}(x,y;g)=1 \ee
On a compact manifold stochastic completeness is always satisfied
\cite{chavel}. Using the properties of heat kernel and stochastic
completeness, one can derive the scaling property of the heat
kernel
\be K_{t}(x,y;g)=\alpha^{D} K_{\alpha^2 t}(x,y;\alpha^{2} g)
\;.\label{heatkernelscaling2} \ee
Free resolvent can be written in terms of the heat kernel
\be R_{0}(x, y |z) = \langle x | \left( H_0 - z \right)^{-1}|y
\rangle = \frac{1}{\hbar}\int_0^{\infty} \mathrm{d} t \;
e^{\frac{z t}{\hbar}} K_t(x,y;g) \;, \label{heatkernel and A} \ee
where $\mathrm{Re}(z)<0$. It can be defined for complex values $z$
by analytic continuation. We have short time asymptotics of the
diagonal heat kernel for any manifold \cite{gilkey}
\be \label{asymheat} \lim_{t\to 0^+} K_{t}(x,x;g) \sim {1\over
(4\pi \hbar t/2m)^{D/2}} \sum_{k=0}^{\infty} u_k(x,x) (\hbar t
/2m)^{k/2} \;, \ee
for every $x\in \mathcal{M}$. Here the functions $u_k(x,x)$ are
scalar polynomials in curvature tensor of the manifold and its
covariant derivatives at the point $x$. When there is no boundary,
the odd terms in the expansion, i.e, $k=1,3,5,...$ vanishes
\cite{kirsten}. In this paper, we always assume that the manifolds
have no boundary. Indeed, we have also short time asymptotic of
the heat kernel for any $x$ and $y$ with the several assumptions
about the structure of the set of geodesics which join the points
$x$ and $y$ \cite{molchanov}. It is shown that (see Theorem 2.1
and 2.2 in \cite{molchanov}) for all $y$ sufficiently close to $x$
(so that $x$ and $y$ can be joined by a unique shortest geodesic
$\gamma_{x,y}$ along which $x$ and $y$ are non-conjugate) then,
\be K_t(x,y;g) \sim {e^{-{m d^2(x,y) \over 2 \hbar t}} \over (4
\pi \hbar t/2m)^{D/2}} d^{(D-1)/2} \Psi_{\gamma}^{-1/2}(x,y) \;,
\ee
where $d(x,y)$ is the geodesic distance between $x$ and $y$ and
$\Psi_{\gamma}(x,y)$ characterizes the divergence of the geodesic
flow near $\gamma$, that is, if we emit a beam of geodesics from
$x$ along $\gamma$ in the solid angle $\mathrm{d}\varphi$
illuminating a hypersurface of area $\mathrm{d}S$ at $y$
orthogonal to $\gamma$, then
$\Psi(x,y)=\mathrm{d}S/\mathrm{d}\varphi$. The function
$\Psi(x,y)$ can also be written in terms of the Jacobi fields
orthogonal to the geodesic $\gamma_{x,y}(s)$, where $0\leq s \leq
d(x,y)$. If the number of shortest geodesics joining $x$ and $y$
is greater than 1, or, if $x$ and $y$ are conjugate along some of
them, then the result takes the following form (up to a bounded
factor)
\be K_t(x,y;g) = \mathcal{O}\left( (\hbar t/2m)^{-{(D+k) \over 2}}
e^{-{m d^2(x,y) \over 2 \hbar t}}\right) \ee
where the index $k=k(x,y)$ depends on the character of the
degeneracy of the geodesic flow between $x$ and $y$. Since all our
calculations essentially give the same physical result for all
cases and subcases \cite{molchanov}, we will just consider a
generic case (Case 3.1 in \cite{molchanov}): The set
$\Omega_{x,y}$ consists of finite number of geodesics
$\gamma_1,...,\gamma_m$ and $x$ and $y$ are non-conjugate along
each of them. In this case, for each $\gamma_i$ we can define
$\Psi_i(x,y) $ by considering the Jacobi fields along $\gamma_i$.
Then, we have (Theorem 3.1 in \cite{molchanov}):
\be K_t(x,y;g) \sim {e^{-{m d^2(x,y) \over 2 \hbar t}} \over (4
\pi \hbar t/2m)^{D/2}} d^{(D-1)/2}(x,y) \sum_i
\Psi_{i}^{-1/2}(x,y) \;. \label{molchanov} \ee

%we will give the result for special compact manifold
%$\mathbb{S}^D$ of radius $R$ for simplicity.
%
%\be K_t(N,S;g) \sim {C_{D-1} R^{D-1} \over \sqrt{\pi} (4 \pi \hbar
%t /2m)^{(2D-1)/2}} \;  e^{-{m(\pi R)^2 \over 2 \hbar t}} \ee
%
%where $C_{D-1}$ is the area of $S^{D-1}$ of radius 1. $N$ and $S$
%are north and south poles, respectively.

\section{Appendix B: Existence of Hamiltonian}
\label{Existence of Hamiltonian}

Let $\Delta$ be a subset of the complex plane. A family $J(E)$, $E
\in \Delta$ of bounded linear operators on the Hilbert space
$\mathcal{H}$ under consideration, which satisfies the resolvent
identity
\be J(E_1)-J(E_2)= (E_1-E_2)J(E_1)J(E_2)\; \label{resolvent
identity}\ee
for $E_1 \; ,E_2 \in \Delta$ is called a pseudo resolvent on
$\Delta$ \cite{pazy}. The following corollary (Corollary 9.5 in
\cite{pazy}) gives the condition for which there exists a densely
defined closed linear operator $A$ such that $J(E)$ is the
resolvent family of $A$: Let $\Delta$ be a unbounded subset of
$\mathbb{C}$ and $J(E)$ be a pseudo resolvent on $\Delta$. If
there is a sequence $E_n \in \Delta$ such that $|E_n| \rightarrow
\infty$ as $n\rightarrow \infty$ and
\be \lim_{n\rightarrow \infty} - E_n J(E_n)x =x \;,
\label{resolvent limit} \ee
for all $x \in \mathcal{H}$, then $J(E)$ is the resolvent of a
unique densely defined closed operator $A$. In order to show the
resolvent kernel that we have found in (\ref{resolventkernel})
corresponds to a unique densely defined closed operator $H$, we
need to prove that it satisfies the resolvent identity, i.e,
\be R(x,y|E_1)-R(x,y|E_2)=(E_1-E_2) \int_{\mathcal{M}}
\mathrm{d}^{D}_{g}z \; R(x,z|E_1)R(z,y|E_2)
\;.\label{resolventkernel identity } \ee
Substituting (\ref{resolventkernel}) into (\ref{resolventkernel
identity }), we obtain
\beqs & & R_0(x,y|E_1)- R_0(x,y|E_2) + \sum_{i,j=1}^{N}
R_0(x,a_i|E_1) \Phi^{-1}_{ij}(E_1) R_0(a_j,y|E_1) \cr \hskip-2cm &
& \hspace{6cm} - \sum_{i,j=1}^{N} R_0(x,a_i|E_2)
\Phi^{-1}_{ij}(E_2) R_0(a_j,y|E_2) \cr & & = (E_1-E_2)
\int_{\mathcal{M}} \mathrm{d}^{D}_{g}z \; \Bigg[
R_0(x,z|E_1)R_0(z,y|E_2) \cr & & \hspace{3cm}+ \sum_{i,j=1}^{N}
R_0(x,z|E_1) R_0(z,a_i|E_2) \Phi^{-1}_{ij}(E_2) R_0(a_j,y|E_2) \cr
& & \hspace{3cm} + \sum_{i,j=1}^{N} R_0(x,a_i|E_1)
\Phi^{-1}_{ij}(E_1) R_0(a_j,z|E_1) R_0(z,y|E_2) \cr & & +
\sum_{i,j=1}^{N}\sum_{k,l=1}^{N} R_0(x,a_i|E_1)
\Phi^{-1}_{ij}(E_1) R_0(a_j,z|E_1) \cr & & \hspace{5cm} \times
R_0(z,a_k|E_2) \Phi^{-1}_{kl}(E_2) R_0(a_l,y|E_2) \Bigg] \;.
\label{resolvent identity prf 1} \eeqs
Using the formula (\ref{heatkernel and A}), it is easy to see that
the free resolvent satisfies the resolvent identity
\beqs & & \hskip-0.6cm (E_1-E_2) \int_{\mathcal{M}}
\mathrm{d}^{D}_{g}z \; R_0(x,z|E_1)R_0(z,y|E_2)\cr & & = (E_1-E_2)
\int_{\mathcal{M}} \mathrm{d}^{D}_{g}z \int_{0}^{\infty}
{\mathrm{d} t_1 \over \hbar} K_{t_1}(x,z;g) e^{t_1 E_1/\hbar}
\int_{0}^{\infty} {\mathrm{d} t_2 \over \hbar} K_{t_2}(z,y;g)
e^{t_2 E_2/\hbar} \cr & & = (E_1-E_2) \int_{0}^{\infty}
\int_{0}^{\infty} {\mathrm{d} t_1 \; \mathrm{d} t_2 \over \hbar^2}
K_{t_1+t_2}(x,y;g)  e^{t_1 E_1/\hbar} e^{t_2 E_2/\hbar} \cr & & =
{(E_1-E_2) \over 2} \int_{0}^{\infty} {\mathrm{d} u \over \hbar}
\Bigg[ \int_{-u}^{u} {\mathrm{d} v \over \hbar} K_{u}(x,y;g)
e^{(u+v)E_1/2\hbar} e^{(u-v)E_2/2\hbar} \Bigg] \cr & & =
\int_{0}^{\infty} {\mathrm{d} u \over \hbar} K_{u}(x,y;g) \left(
e^{u E_1/\hbar}- e^{u E_2/\hbar} \right) =
R_0(x,y|E_1)-R_0(x,y|E_2) \;, \label{free resolvent identity}
\eeqs
where we have used the semigroup property of heat kernel
(\ref{semigroupprop}) and made the change of variables
$u=t_1+t_2$, $v=t_1-t_2$. Then, the equation (\ref{resolvent
identity prf 1}) becomes
\beqs & & \sum_{i,j=1}^{N} R_0(x,a_i|E_1) \Phi^{-1}_{ij}(E_1)
R_0(a_j,y|E_1) \cr \hskip-2cm & & \hspace{6cm} - \sum_{i,j=1}^{N}
R_0(x,a_i|E_2) \Phi^{-1}_{ij}(E_2) R_0(a_j,y|E_2) \cr & & =
(E_1-E_2) \int_{\mathcal{M}} \mathrm{d}^{D}_{g}z \; \Bigg[
\sum_{i,j=1}^{N} R_0(x,z|E_1) R_0(z,a_i|E_2) \Phi^{-1}_{ij}(E_2)
R_0(a_j,y|E_2) \cr & & \hspace{3cm} + \sum_{i,j=1}^{N}
R_0(x,a_i|E_1) \Phi^{-1}_{ij}(E_1) R_0(a_j,z|E_1) R_0(z,y|E_2) \cr
& & + \sum_{i,j=1}^{N}\sum_{k,l=1}^{N} R_0(x,a_i|E_1)
\Phi^{-1}_{ij}(E_1) R_0(a_j,z|E_1) \cr & & \hspace{5cm} \times
R_0(z,a_k|E_2) \Phi^{-1}_{kl}(E_2) R_0(a_l,y|E_2) \Bigg] \;.
\label{resolvent identity prf 2} \eeqs
If we add and subtract the terms $\sum_{i,j=1}^{N} R_0(x,a_i|E_1)
\Phi^{-1}_{ij}(E_1) R_0(a_j,y|E_2)$, \\
$\sum_{i,j=1}^{N} R_0(x,a_i|E_1) \Phi^{-1}_{ij}(E_2)
R_0(a_j,y|E_2)$ on the left hand side of (\ref{resolvent identity
prf 2}), and rearrange all the terms, we obtain
\beqs & & \sum_{i,j=1}^{N} R_0(x,a_i|E_1) \Phi^{-1}_{ij}(E_1)
\left[ R_0(a_j,y|E_1)- R_0(a_j,y|E_2)\right] \cr & & +
\sum_{i,j=1}^{N} R_0(x,a_i|E_1)
\left[\Phi^{-1}_{ij}(E_1)-\Phi^{-1}_{ij}(E_2)\right]
R_0(a_j,y|E_2) \cr & & + \sum_{i,j=1}^{N}
\left[R_0(x,a_i|E_1)-R_0(x,a_i|E_2)\right] \Phi^{-1}_{ij}(E_2)
R_0(a_j,y|E_2) \cr & & = (E_1-E_2) \sum_{i,j=1}^{N} R_0(x,a_i|E_1)
\Phi^{-1}_{ij}(E_1) \int_{\mathcal{M}} \mathrm{d}^{D}_{g} z \;
R_0(a_j,z|E_1)R_0(z,y|E_2) \cr & & + \sum_{i,j=1}^{N}
R_0(x,a_i|E_1)
\left[\Phi^{-1}_{ij}(E_1)-\Phi^{-1}_{ij}(E_2)\right]
R_0(a_j,y|E_2) \cr & & + (E_1-E_2) \sum_{i,j=1}^{N}
\int_{\mathcal{M}} \mathrm{d}^{D}_{g} z \;
R_0(x,z|E_1)R_0(z,a_i|E_2) \Phi^{-1}_{ij}(E_2) R_0(a_j,y|E_2)
\label{resolvent identity prf 3} \;,\eeqs
where we have used the result (\ref{free resolvent identity}) in
the first and third terms. The second term can be written as
\beqs & & \sum_{i,j=1}^{N} \sum_{k,l=1}^{N} R_0(x,a_i|E_1)
\Phi^{-1}_{ik}(E_1)
\left[\Phi_{kl}(E_2)-\Phi_{kl}(E_1)\right]\Phi^{-1}_{lj}(E_2)
R_0(a_j,y|E_2) \;. \label{resolvent 2nd term s1} \eeqs
It is important to notice that difference of the principal matrix
equals to the difference in free resolvent kernel, that is,
\beqs & &
\Phi_{ij}(E_2)-\Phi_{ij}(E_1)=R_0(a_i,a_j|E_1)-R_0(a_i,a_j|E_2)
\cr & & = (E_1-E_2)\int_{\mathcal{M}} \mathrm{d}^{D}_{g}z \;
R_0(a_i,z|E_1)R_0(z,a_j|E_2) \;, \label{resolvent 2nd term} \eeqs
for any $i$ and $j$. After substituting (\ref{resolvent 2nd term})
into (\ref{resolvent 2nd term s1}), the equation (\ref{resolvent
identity prf 3}) becomes
\beqs & & (E_1-E_2) \sum_{i,j=1}^{N} R_0(x,a_i|E_1)
\Phi^{-1}_{ij}(E_1) \int_{\mathcal{M}} \mathrm{d}^{D}_{g} z \;
R_0(a_j,z|E_1)R_0(z,y|E_2) \cr & & + (E_1-E_2) \sum_{i,j=1}^{N}
\sum_{k,l=1}^{N} R_0(x,a_i|E_1) \Phi^{-1}_{ij}(E_1) \cr & &
\hspace{3cm} \times \int_{\mathcal{M}} \mathrm{d}^{D}_{g}z \;
R_0(a_j,z|E_1)R_0(z,a_k|E_2) \Phi^{-1}_{kl}(E_2) R_0(a_l,y|E_2)
 \cr & & + (E_1-E_2)
\sum_{i,j=1}^{N} \int_{\mathcal{M}} \mathrm{d}^{D}_{g} z \;
R_0(x,z|E_1)R_0(z,a_i|E_2) \Phi^{-1}_{ij}(E_2) R_0(a_j,y|E_2)
\label{resolvent identity prf 4} \;,\eeqs
This is exactly equal to (\ref{resolvent identity prf 2}) so
resolvent identity is satisfied. We must now imply the following
condition in $L^2$ norm
\be ||E_n R(E_n)f+f|| \rightarrow 0 \;, \ee
as $n\rightarrow \infty$ and $f$ belongs to the Hilbert space
$\mathcal{H}$ under consideration and the norm is taken with
respect to $\mathcal{H}$. Let us choose the sequence $E_n = -n
E_0$ since the resolvent is well defined in this resolvent set, in
which we have no spectrum below the absolute value of the bound
$E_0$ that we have found for the ground state energy. Without loss
of generality, we can set $E_0=c|E_*|$, where $c>2$. Then, we have
\be || n E_0 R(-nE_0)f - f || \rightarrow 0 \;, \ee
as $n\rightarrow \infty$. Using (\ref{resolventkernel}) and
separating the free part, we get
\beqs  || n E_0 R(-nE_0)f -f || & \leq & || n E_0 R_0(-nE_0)f -f
|| \cr & +& n E_0 || R_0(-nE_0) \Phi^{-1}(-n E_0) R_0(-nE_0) f ||
\;. \eeqs
Since it is well known that the first part of the sum converges to
zero as $n\rightarrow \infty$, that is, free resolvent defines a
densely defined closed operator (Laplacian), we are going to
investigate only the second term
\beqs  & &  n E_0 || R_0(-nE_0) \Phi^{-1}(-n E_0) R_0(-nE_0) f ||
\cr & & \leq n E_0 \Bigg[ \sum_{i,j,k,l=1}^{N} \int_{\mathcal{M}}
\mathrm{d}^{D}_{g} x \; R_0(a_i,x|-n E_0) R_0(x,a_l|-n E_0) \cr &
& \times \int_{\mathcal{M}} \mathrm{d}^{D}_{g} y \; R_0(a_j,y|-n
E_0) R_0(y,a_k|-n E_0)
 |\Phi_{ij}^{-1}(-n E_0)| |\Phi_{kl}^{-1}(-n E_0)|\Bigg]^{1/2} \;, \label{resolvent inequality} \eeqs
where we have used the fact that the Hilbert space norm of an
operator is smaller than its Hilbert-Schmidt norm: $||A f|| \leq
\mathrm{Tr}^{1/2}(A^{\dagger} A)$ with $A=R_0(-nE_0) \Phi^{-1}(-n E_0) \\
R_0(-nE_0)$. By using (\ref{heatkernel and A}) and the change of
variables $u=t_1+ t_2$, $v=t_1-t_2$, we get
\beqs & & \hskip-1cm  \int_{\mathcal{M}} \mathrm{d}^{D}_{g} x \;
R_0(a_i,x|-n E_0) R_0(x,a_l|-n E_0) =
\int_{0}^{\infty}\int_{0}^{\infty} {\mathrm{d} t_1 \, \mathrm{d}
t_2 \over \hbar^2} \; K_{t_1 + t_2} (a_i,a_l;g) e^{-n {(t_1 +
t_2)E_0 \over \hbar}} \cr & &  = \int_{0}^{\infty} {\mathrm{d} t
\over \hbar^2} \; t \; K_{t} (a_i,a_l;g) e^{-n {t E_0 \over
\hbar}} \label{rzero2 heat kernel} \eeqs
Let us first consider the diagonal case $i=l$ and $k=j$ in the
above. Then, the equation (\ref{resolvent inequality}) becomes
\beqs  & & n E_0 \Bigg[ \sum_{i,j=1}^{N} \int_{\mathcal{M}}
\mathrm{d}^{D}_{g} x \; R_0(a_i,x|-n E_0) R_0(x,a_i|-n E_0) \cr &
& \times \int_{\mathcal{M}} \mathrm{d}^{D}_{g} y \; R_0(a_j,y|-n
E_0) R_0(y,a_j|-n E_0)
 |\Phi_{ij}^{-1}(-n E_0)| |\Phi_{ji}^{-1}(-n E_0)|\Bigg]^{1/2} \;. \label{norm bound} \eeqs
In this case, the upper bound of the equation (\ref{rzero2 heat
kernel}) can be found from the upper bound of the heat kernel for
compact manifolds
\beqs & & \hskip-1cm  \int_{\mathcal{M}} \mathrm{d}^{D}_{g} x \;
R_0(a_i,x|-n E_0) R_0(x,a_l|-n E_0) \cr & & \leq {4A \over
V(\mathcal{M})} {1 \over E_{0}^{2}n^2} +  {4A B(\varepsilon) \over
\hbar^2 (\hbar/2m)^{D/2}} \left({n E_0 \over
\hbar}\right)^{{D\over 2}-2} \Gamma\left(2-{D \over 2}\right) \;,
\label{r02 compact upper bound} \eeqs
and for Cartan-Hadamard manifolds
\beqs & & \hskip-1cm \int_{\mathcal{M}} \mathrm{d}^{D}_{g} x \;
R_0(a_i,x|-n E_0) R_0(x,a_l|-n E_0) \cr & & \leq
{C(\varepsilon,\kappa) \over \hbar^2 (4\pi \hbar/2m)^{D/2}}
\left({n E_0 \over \hbar}\right)^{{D\over 2}-2} \Gamma\left(2-{D
\over 2}\right) \;. \label{r02 CH upper bound} \eeqs
In order to give the upper bound for the inverse principal matrix,
we decompose the principal matrix into two positive matrices
\be \Phi = D - K \ee
where $D$ and $K$ stand for the diagonal and off diagonal parts of
the principal matrix. Then, it is easy to see $\Phi =
D(\mathbf{1}-D^{-1}K)$. The principal matrix is invertible if and
only if $(\mathbf{1}-D^{-1}K)$, and $(\mathbf{1}-D^{-1}K)$ has an
inverse if the norm $||D^{-1}K||<1$. Then, we write the inverse of
$\Phi$ as a geometric series
\beqs \Phi^{-1}&=&(\mathbf{1}-D^{-1}K)^{-1} D^{-1} \cr & = &
\left(1 + (D^{-1}K) + (D^{-1}K)^2+ ... \right) D^{-1} \;,\eeqs
where we must have $||D^{-1}K||< 1$. Since we are not concerned
with the sharp bounds on $\Phi^{-1}$ for this problem, we can
choose $||D^{-1}K||< 1/2$ by adjusting the $n E_0$ sufficiently
large without loss of generality and get
\be |\Phi^{-1}| \leq 2 |D^{-1}| \;. \ee
The lower bound of the diagonal principal matrix for compact
(\ref{lower bound diagonal phi compact}) and Cartan-Hadamard
manifolds (\ref{lower bound diagonal phi CH}) gives the upper
bound of the inverse principal matrix. Hence, we find
\be \label{compact phi upper} |\Phi_{ii}^{-1} (-n E_0)| \leq
\begin{cases}
\begin{split}
\left(4 \pi \hbar^2 /2m \right) \ln^{-1} \left( n E_0 /\mu^2
\right)
\end{split}
& \textrm{if $D = 2$} \\
\begin{split}
{\hbar \left(4 \pi \hbar /2m \right)^{3/2} \over 2\sqrt{\pi}}
\left( \sqrt{{n E_0 \over \hbar}} - \sqrt{{\mu^2 \over \hbar}}
\right)^{-1}
\end{split}
& \textrm{if $D = 3$}\;,
\end{cases}
\ee
for compact manifolds and

\be \label{cartanhadamard phi upper} | \Phi_{ii}^{-1}(-n E_0)|
\leq
\begin{cases}
\begin{split}
{\left(4 \pi \hbar^2 /2m \right) \over c} \ln^{-1} \left( { {n E_0
\over \hbar} + \xi \over {\mu^2 \over \hbar} + \xi } \right)
\end{split}
& \textrm{if $D = 2$} \\
\begin{split}
{\hbar \left(4 \pi \hbar /2m \right)^{3/2} \over 2\sqrt{\pi} c}
\left(\sqrt{{n E_0 \over \hbar} + \xi} - \sqrt{{\mu^2 \over \hbar}
+\xi} \right)^{-1}
\end{split}
& \textrm{if $D = 3$} \;,
\end{cases}
\ee
for Cartan-Hadamard manifolds. If we substitute the results
(\ref{r02 compact upper bound}) and (\ref{r02 CH upper bound})
into (\ref{norm bound}) for $D=2$, and take the limit
$n\rightarrow \infty$, the result goes to zero. Since the norm is
always  positive, we prove
\be || n E_0 R(-nE_0)f -f || \rightarrow 0 \ee
as $n\rightarrow \infty$. It is almost evident that the off
diagonal terms in the sum also vanishes in this limit primarily
because these terms are exponentially damped $e^{-\sqrt{n}}$ due
to the upper bound of the heat kernel. Unfortunately, the proof
for $D=3$ is more subtle and the volume growth conditions of the
manifolds are very delicate in the analysis so we can not prove it
for three dimensional case by the same approach. Instead we follow
the following approach: In three dimensions,  estimating the
Hilbert space norm by  Hilbert-Schmidt norm does not lead to zero.
Instead we will first show that the last term
\beqs &\ & |E_n|  \Bigg[\int_{\cal M} \mathrm{d}_g^3 x \;
\sum_{i,j,k,l=1}^{N} R_0(x,a_i|E_n)\Phi^{-1}_{ij}(E_n) \int_{\cal
M} \mathrm{d}_g^3 z \;  R_0(a_j, z|E_n) f^*(z)  \cr
  &\ & \ \ \ \ \   \ \ \ \ \ \ \ \ R_0(x, a_k|E_n)
  \Phi^{-1}_{kl}(E_n)\int_{\cal M} \mathrm{d}_g^3 y \; R_0(a_l,y|E_n)f(y)\Bigg]^{1/2}
\label{three ham ex} \; \eeqs
goes to zero as $E_n \to -\infty$ for any $f\in L^2(\mathcal{M})$.
From our previous argument, we know that the inverse of the
principal matrix $\Phi$ satisfies:
\be {\rm max}_{ij} \ |\Phi^{-1}_{ij} (E_n)| \leq  {A_2 \over
|E_n|^{1/2}} \;, \ee
where we define all the constant terms coming from the bounds of
the heat kernel as $A_2$ (exact form of the constants is not
important here) and ignore the term in the denominator for large
values of $n$ for simplicity, which can be read from (\ref{compact
phi upper}) and (\ref{cartanhadamard phi upper}). We shall use the
notation for the constants coming from the bounds of the heat
kernel combined with the other constants factors as
$A_1,A_2,A_3,\ldots$ for simplicity. Moreover, we can combine the
two resolvents with the common variable $x$, and as a result, we
can express this combination as
\be \Bigg[\int_0^\infty \ {\mathrm{d} t \over \hbar}  \; {t \over
\hbar} e^{-{|E_n|t \over \hbar}} K_t(a_i, a_k;g) \Bigg]^{1/2} ,\ee
and pull it out of the square root. Using similar arguments as
before, we can show that this term in three dimensions, for both
Cartan-Hadamard type manifolds and Compact manifolds (bounded
Ricci),  including the identical  beginning and end points, is
smaller than
\be  {A_3 \over |E_n|^{1/4}} \;, \ee
where $A_3$ can easily be read from the upper bound of the heat
kernel. Hence we end up with the fact that the expression
(\ref{three ham ex}) is smaller than,

\be N A_4 |E_n|^{1/4} \sum_{j,l=1}^{N} \Bigg[\int_{\cal M}
\mathrm{d}_g^3 y \; R_0(a_j, z|E_n)|f(z)| \int_{\cal M}
\mathrm{d}_g^3 z \; R_0( y, a_l)|f(y)|\Bigg]^{1/2} .\ee
Hence we should show that the term
\be \int_{\cal M} \mathrm{d}_g^3 y \; R_0(a_j, y|E_n)|f(y)| \ee
decays faster than $|E_n|^{-1/4}$.

To do this,  we will pick any one of the centers and choose
Riemannian normal coordinates around it, we assume that the
injectivity radius of the manifold is $\delta>0$. But first we
reexpress this term in terms of the heat kernel and use some
bounds,
\beqs R_0(a_j, y|E_n)&=& \; \int_0^\infty {\mathrm{d} t \over
\hbar} e^{-{|E_n|t \over \hbar}} K_t(a_j,y;g) \cr \nonumber & \leq
& {A_5} \int_0^\infty \mathrm{d} t \; {1\over t^{3/2}} e^{-{m
d^2(a_j,y)\over \hbar C_2 t}-{|E_n|t \over \hbar}} ,\eeqs
for Cartan-Hadamard manifolds, and for compact manifolds we have a
similar term with an inverse volume term, ${1\over V({\cal M})}$,
added. In the case of compact manifolds volume contribution term
goes to zero faster than $|E_n|^{-1/4}$ as can be checked easily,
so it causes no problems. In both cases we will concentrate on the
least convergent part. If we evaluate the integral over $t$ now we
find,
\be \int_{\cal M}  \mathrm{d}_g^3 y \; R_0(a_j, y|E_n)|f(y)| \leq
A_6 \int_{\cal M} \mathrm{d}_g^3 y \; e^{-2 \sqrt{{m d^2(a_j,y)
|E_n| \over \hbar^2 C_2}}} {|f(y)|\over d(a_j,y)} \;. \ee
We divide the right hand side as,
\be \int_{B_\delta(a_j)} \mathrm{d}_g^3 y \; e^{-2 \sqrt{{m
d^2(a_j,y) |E_n| \over \hbar^2 C_2}}} {|f(y)|\over d(a_j,y)}
+\int_{{\cal{M}}\setminus B_\delta(a_j)} \mathrm{d}_g^3 y \;e^{-2
\sqrt{{m d^2(a_j,y) |E_n| \over \hbar^2 C_2}}} {|f(y)|\over
d(a_j,y)} \;; \ee
here the last term is smaller than
\beqs &\ & {e^{-\sqrt{{m \delta^2 |E_n| \over \hbar^2 C_2}}} \over
\delta} \int_{{\cal{M}}\setminus B_\delta(a_j)} \mathrm{d}_g^3 y
\; e^{-\sqrt{{m d^2(a_j,y) |E_n| \over \hbar^2 C_2}}}
|f(y)|\cr\nonumber &\ & \ \ \leq {e^{-\sqrt{{m \delta^2 |E_n|
\over \hbar^2 C_2}}} \over \delta} \int_{{\cal{M}}} \mathrm{d}_g^3
y \; e^{-\sqrt{{m d^2(a_j,y) |E_n| \over \hbar^2 C_2}}} |f(y)| \cr
& & \leq {e^{-\sqrt{{m \delta^2 |E_n| \over \hbar^2 C_2}}} \over
\delta} \Bigg[\int_{\cal{M}} \mathrm{d}_g^3 y \; e^{-2 \sqrt{{m
d^2(a_j,y) |E_n| \over \hbar^2 C_2}}} \Bigg]^{1/2}||f||_2 \eeqs
By a theorem of Gaffney \cite{gaffney}, for a stochastically
complete manifold, for any $\alpha>0$ (with the inverse length
dimension) and any point $a$ on the manifold, the integrals
satisfy,
\be \int_{\cal M} \mathrm{d}_g^3 y \; e^{-\alpha d(a,y)} \leq
\infty \;,\ee
This establishes that the last term decays faster than
$|E_n|^{-1/4}$, so we should look at the first part. As has been
said we go to the Riemann normal coordinates of the geodesic ball
of radius $\delta$ and write the integral in terms of Gaussian
spherical representation:
\be
 \int_{\mathbb{S}^2} \mathrm{d}\Omega \int_0^\delta \mathrm{d}r \; r^2 J(r,\theta) e^{-2 \sqrt{{m r^2
|E_n| \over \hbar^2 C_2}}} {|f(r,\theta)|\over r} \;. \ee
Let us recall that in Gaussian spherical coordinates, the integral
of a function $f$ on an $D$-dimensional Riemannian manifold
$\mathcal{M}$ becomes,
\be \int_{\cal M} \mathrm{d}_g^D x \; f(x) =
\int_{\mathbb{S}^{D-1}} \mathrm{d}\Omega \int_0^{\rho_\Omega}
\mathrm{d}r \; r^{D-1} f(r,\theta) J(r,\theta). \ee
Here $\Omega$ denotes the direction in the tangent space around a
point that we choose, and $\rho_\Omega$ refers to distance to the
cut locus of the point in the direction $\Omega$. We will now
divide the integral over $r$ to two parts,
\be \int_{\mathbb{S}^2} \mathrm{d}\Omega \int_0^\Delta \mathrm{d}
r \; r^2 J(r,\theta) e^{-2 \sqrt{{m r^2 |E_n| \over \hbar^2 C_2}}}
{|f(r,\theta)|\over r}+ \int_{\mathbb{S}^2} \mathrm{d} \Omega
\int_\Delta^\delta \mathrm{d} r \; r^2 J(r,\theta) e^{-2 \sqrt{{m
r^2 |E_n| \over \hbar^2 C_2}}} {|f(r,\theta)|\over r} \;, \ee
where $0 < \Delta < \delta$. Let us now introduce the following
function:
\be \textrm{sn}_k(r)=
\begin{cases}
{\sin(\sqrt{k} r) \over \sqrt{k}}& {\rm if} \ k>0\\
r & {\rm if} \ k=0\\
{\sinh(\sqrt{-k} r) \over \sqrt{-k}}& {\rm if} \ k>0\\
\end{cases}
\ee
This function is very useful for the Bishop-Gunther volume
comparison theorems. We assume that ${\cal M}$ has Ricci tensor
bounded from below by $k_1$, i.e. $ {\rm Ric}(.,.)> k_1g(.,.)$ and
sectional curvature $K$ bounded from above by $k_2$. Then, the
Jacobian factor of the Gaussian spherical coordinates satisfies an
inequality as follows \cite{gallot},
\be {\textrm{sn}^2_{k_2}(r)\over r^2}< J(r,\theta)<
{\textrm{sn}^2_{k_1}(r)\over r^2} .\ee
In the second integral, we use
\be \begin{split} \int_\Delta^\delta & \mathrm{d} r \; r \; e^{-2
\sqrt{{m r^2 |E_n| \over \hbar^2 C_2}}} \int_{\mathbb{S}^2}
\mathrm{d} \Omega |f(r,\theta)|J^{1/2}(r,\theta) \;
J^{1/2}(r,\theta)
\\ & \leq \Bigg[\int_\Delta^\delta \mathrm{d} r \; r^2 \Bigg(\int_{\mathbb{S}^2}
\mathrm{d} \Omega |f(r,\theta)|J^{1/2}(r,\theta) \Bigg)^2
\Bigg]^{1/2}\Bigg[\int_\Delta^\delta \mathrm{d} r \; {{\rm
sn}_{k_1}^2(r)\over r^2}e^{-4 \sqrt{{m r^2 |E_n| \over \hbar^2
C_2}}} \Bigg]^{1/2} \end{split} \ee
where we use the Bishop-Gunther  volume comparison theorem again,
for $J$. Let us now make the  observation that there are constants
$A_+,A_-$, which depend only on $\delta$ and  $k_i$'s such that,
\be
 A_-(k_i,k_j)<{{\rm sn}_{k_i}(r)\over {\rm sn}_{k_j}(r)}<A_+(k_i,k_j)
\ee
for $r\in [0,\delta]$.  This  can now be invoked at the second
piece, giving us,
\be \begin{split} \int_\Delta^\delta & \mathrm{d} r \; r \; e^{-2
\sqrt{{m r^2 |E_n| \over \hbar^2 C_2}}} \int_{\mathbb{S}^2}
\mathrm{d} \Omega |f(r,\theta)|J^{1/2}(r,\theta) \;
J^{1/2}(r,\theta) \\ & \leq \Bigg[\Bigg(\int_\Delta^\delta
\mathrm{d}r \;  r^2 \int_{\mathbb{S}^2} \mathrm{d}\Omega
|f(r,\theta)|^2 J(r,\theta) \Bigg)\Bigg(\int_{\mathbb{S}^2}
\mathrm{d} \Omega\Bigg)
\Bigg]^{1/2}A_+(k_1,0)\Bigg[\int_\Delta^\delta \mathrm{d} r e^{-4
\sqrt{{m r^2 |E_n| \over \hbar^2 C_2}}} \Bigg]^{1/2}\\ & \leq ||f||_2
(4\pi)^{1/2} A_+(k_1,0){1\over 2(m/\hbar^2 C_2)^{1/4} |E_n|^{1/4}}
e^{-2 \sqrt{{m \Delta^2 |E_n| \over \hbar^2 C_2}}}
\end{split} \ee
If we choose $\Delta=(\hbar^2 R/m)^{1/3}|E_n|^{-1/3}$, the
exponent goes to zero as $|E_n|\to \infty$. For the first part of
the integral we use the following characterization of essential
supremum: let us define
\be
 \Lambda(\epsilon)=\mu(\{r\in [0, \Delta]| \ |r^{3/2}F(r)|>\epsilon\})
 ,\ee
then we have
\be
 \essup_{[0, \Delta]}|r^{3/2}F(r)|=\inf_\epsilon \{\epsilon| \Lambda(\epsilon)=0\}
\;. \ee
Let us use now $F(r)=\int_{\mathbb{S}^2} \mathrm{d}\Omega
f(r,\theta)$, and using Bishop-Gunther bound  for the first part
as,
\be \int_0^\Delta \mathrm{d} r \; r^{3/2} \int_{\mathbb{S}^2}
\mathrm{d}\Omega \; |f(r,\theta)| {e^{-2 \sqrt{{m r^2 |E_n| \over
\hbar^2 C_2}}} \over r^{1/2}} {\textrm{sn}^2_{k_1}(r)\over r^2}
\ee
which is smaller than;
\beqs &\ & A_+^2(k_1,0)\int_0^\Delta \mathrm{d} r \; r^{3/2}
\int_{\mathbb{S}^2} \mathrm{d}\Omega \; |f(r,\theta)| {e^{-2
\sqrt{{m r^2 |E_n| \over \hbar^2 C_2}}} \over r^{1/2}}\cr\nonumber
&\ & \ \ \ \ \ \leq
A_+^2(k_1,0)\Big(\essup_{[0,\Delta]}|r^{3/2}F(r)|\Big)\Bigg(
\int_0^\Delta \mathrm{d} r \; {e^{-2 \sqrt{{m r^2 |E_n| \over
\hbar^2 C_2}}} \over r^{1/2}} \Bigg)\cr\nonumber &\ & \ \ \ \ \
\leq A_+^2(k_1,0)\Big(\essup_{[0,\Delta]}|r^{3/2}F(r)|\Big) {1\over
2(m/\hbar^2 C_2)^{1/4} |E_n|^{1/4}} \eeqs
If we know take the limit $\Delta=(\hbar^2/
mR^2)^{1/3}|E_n|^{-1/3}\to 0$, we claim that the
essential-suppremum goes to zero. To see this observe by Markov
inequality \cite{stein} that
\beqs \Lambda(\epsilon)& \leq & {1\over \epsilon} \int_0^\Delta
\mathrm{d} r \; |r^{3/2}F(r)|\cr \nonumber & \leq & {1\over
\epsilon}\Bigg( \int_0^\Delta \mathrm{d}r \; r
\Bigg)^{1/2}\Bigg(\int_0^\Delta \mathrm{d}r \; r^2
\left(\int_{\mathbb{S}^2} \mathrm{d}\Omega \;
|f(r,\theta)|\right)^2 \Bigg)^{1/2}\cr\nonumber & \leq & {1\over
\epsilon}{\Delta\over \sqrt{2}}\Bigg(\int_0^\Delta \mathrm{d} r \;
{r^2\over \textrm{sn}^2_{k_2}(r)}
\textrm{sn}^2_{k_2}(r)\int_{\mathbb{S}^2} \mathrm{d} \Omega \;
|f(r,\theta)|^2 \int_{\mathbb{S}^2}\mathrm{d} \Omega \Bigg)^{1/2}
\cr\nonumber & \leq & {1\over \epsilon}{\Delta\over
\sqrt{2}}(4\pi)^{1/2} A_+(0, k_2)\Bigg(\int_0^\Delta \mathrm{d} r
\;  \textrm{sn}^2_{k_2}(r)\int_{\mathbb{S}^2} \mathrm{d} \Omega \;
|f(r,\theta)|^2\Bigg)^{1/2} \cr\nonumber & \leq & {1\over
\epsilon}{\Delta\over \sqrt{2}}(4\pi)^{1/2} A_+(0,
k_2)\Bigg(\int_0^\Delta \mathrm{d} r \; r^2 \int_{\mathbb{S}^2}
\mathrm{d}\Omega \; J(r,\theta) |f(r,\theta)|^2\Bigg)^{1/2}
\cr\nonumber
 & \leq &  {1\over \epsilon}{\Delta\over \sqrt{2}}(4\pi)^{1/2} A_+(0, k_2)||f||_2
.\eeqs
 hence for any $\epsilon>0$, as $\Delta\to 0$ we can make $\Lambda(\epsilon)=0$, thus the infimum goes to zero in this
 limit. As a result we see that the equation (\ref{three ham ex}) is
 smaller than
 \be \begin{split}
 & {A_+^2(k_1,0) \over 2(m/\hbar^2 C_2)^{1/4}} \Big(\essup_{[0,\Delta]}|r^{3/2}F(r)|\Big) +||f||_2 (4\pi)^{1/2} {A_+
 (k_1,0) \over 2(m/\hbar^2 C_2)^{1/4}} e^{-|E_n|^{1/6}(m R^2/\hbar^2)^{1/6}} \\ &
  +
{e^{-\sqrt{{m \delta^2 |E_n| \over \hbar^2 C_2}}} \over \delta}
\Bigg[\int_{\cal{M}} \mathrm{d}_g^3 y \; e^{-2 \sqrt{{m d^2(a_j,y)
|E_n| \over \hbar^2 C_2}}} \Bigg]^{1/2}||f||_2
  \to 0 \ \
{\rm as} \ \ |E_n|\to \infty \;, \end{split} \ee and as a result
goes to zero as desired. This completes the proof of the existence
of the Hamiltonian in three dimensions.

\end{document}